\documentclass[10pt,journal]{IEEEtran}		
\usepackage[T1]{fontenc} 
\usepackage{bm} 
\usepackage{setspace} 
\usepackage
[bookmarks=true,
colorlinks=true,
linkcolor=black,
anchorcolor=black,
urlcolor=black,
citecolor=black]{hyperref}
\usepackage{cite}
\usepackage{float}
\usepackage{graphicx}
\usepackage{caption}
\usepackage{subcaption}
\usepackage{tabularx}
\usepackage{booktabs}
\usepackage[table]{xcolor}
\newcolumntype{L}{>{\hspace*{-\tabcolsep}}l}
\newcolumntype{R}{c<{\hspace*{-\tabcolsep}}}
\definecolor{lightblue}{rgb}{0.93,0.95,1.0}
\usepackage{ifthen}
\usepackage{amsmath}
\usepackage{amssymb}
\usepackage{amsfonts}
\usepackage{commath}
\allowdisplaybreaks[4]
\usepackage{indentfirst}
\usepackage{algorithm,algorithmic}

\usepackage{cleveref}

\newtheorem{mytheorem}{Theorem}

\newcommand{\diag}{\mathrm{diag}}
\newcommand{\rank}{\mathrm{rank}}
\newcommand{\spann}{\mathrm{span}}

\newcommand{\trace}{\mathrm{tr}}

\newcommand{\LowSNR}{\mathrm{LS}}
\newcommand{\HighRiceK}{\mathrm{HR}}
\newcommand{\LoS}{\mathrm{LoS}}
\newcommand{\NLoS}{\mathrm{NLoS}}

\newcommand{\opt}{\mathrm{opt}}

\newcommand{\syn}{\mathrm{cps}}
\newcommand{\SUM}{\mathrm{sum}}

\newcommand{\dint}{\,\mathrm{d}}

\newcommand{\ut}{\mathrm{ut}}
\newcommand{\sat}{\mathrm{sat}}

\newcommand{\Sone}{\tdS}
\newcommand{\Mx}{M_{\rx}}
\newcommand{\My}{M_{\ry}}

\newcommand{\Nx}{N_{\rx'}}
\newcommand{\Ny}{N_{\ry'}}

\newcommand{\nv}{n_{\rv}}
\newcommand{\Niter}{N_{\mathrm{iter}}}

\newcommand{\Nsc}{N_{\mathrm{sc}}}
\newcommand{\Ncp}{N_{\mathrm{cp}}}
\newcommand{\Tsc}{T_{\mathrm{sc}}}
\newcommand{\Tcp}{T_{\mathrm{cp}}}
\newcommand{\Ts}{T_{\mathrm{s}}}
\newcommand{\Complex}[2]{\bbC^{#1 \times #2}}
\newcommand{\Real}[2]{\bbR^{#1 \times #2}}

\newcommand{\Xinv}[1]{\left(#1\right)^{-1}}
\newcommand{\Xginv}[1]{\left(#1\right)^{-}}

\newcommand{\xinv}[1]{\frac{1}{#1}}

\newcommand{\xdeg}[1]{#1^{\circ}}
\newcommand{\sabs}[1]{\lvert#1\rvert}
\newcommand{\snorm}[1]{\lVert#1\rVert}

\newcommand{\bbR}{\mathbb{R}}
\newcommand{\bbC}{\mathbb{C}}
\newcommand{\bbE}{\mathbb{E}}

\newcommand{\clD}{\mathcal{D}}

\newcommand{\clK}{\mathcal{K}}
\newcommand{\clL}{\mathcal{L}}

\newcommand{\clN}{\mathcal{N}}

\newcommand{\clO}{\mathcal{O}}

\newcommand{\clW}{\mathcal{W}}

\newcommand{\clCN}{\mathcal{CN}}

\newcommand{\re}{\mathrm{e}}
\newcommand{\rf}{\mathrm{f}}

\newcommand{\rn}{\mathrm{n}}

\newcommand{\rx}{\mathrm{x}}
\newcommand{\ry}{\mathrm{y}}
\newcommand{\rz}{\mathrm{z}}

\newcommand{\rv}{\mathrm{v}}

\newcommand{\rB}{\mathrm{B}}

\newcommand{\rQ}{\mathrm{Q}}
\newcommand{\rR}{\mathrm{R}}

\newcommand{\rT}{\mathrm{T}}

\newcommand{\vphi}{\varphi}

\newcommand{\ckbdH}{\check{\bdH}}

\newcommand{\brbdH}{\bar{\bdH}}

\newcommand{\tdc}{\tilde{c}}

\newcommand{\tdS}{\tilde{S}}

\newcommand{\tdbdc}{\tilde{\bdc}}

\newcommand{\tdbdd}{\tilde{\bdd}}

\newcommand{\tdbdA}{\tilde{\bdA}}

\newcommand{\tdbdT}{\tilde{\bdT}}

\newcommand{\udC}{\underline{C}}
\newcommand{\udR}{\underline{R}}

\newcommand{\udalpha}{\underline{\alpha}}

\newcommand{\udbdu}{\underline{\bdu}}

\newcommand{\udbdm}{\underline{\bdm}}

\newcommand{\udbdM}{\underline{\bdM}}

\newcommand{\udbdT}{\underline{\bdT}}

\newcommand{\udtdbdc}{\underline{\tdbdc}}

\newcommand{\bdzro}{\mathbf{0}}
\newcommand{\bdone}{\mathbf{1}}
\newcommand{\bda}{\mathbf{a}}
\newcommand{\bdb}{\mathbf{b}}
\newcommand{\bdc}{\mathbf{c}}
\newcommand{\bdd}{\mathbf{d}}

\newcommand{\bdg}{\mathbf{g}}

\newcommand{\bdm}{\mathbf{m}}

\newcommand{\bdp}{\mathbf{p}}

\newcommand{\bds}{\mathbf{s}}

\newcommand{\bdu}{\mathbf{u}}

\newcommand{\bdw}{\mathbf{w}}
\newcommand{\bdx}{\mathbf{x}}
\newcommand{\bdy}{\mathbf{y}}
\newcommand{\bdz}{\mathbf{z}}
\newcommand{\bdA}{\mathbf{A}}
\newcommand{\bdB}{\mathbf{B}}
\newcommand{\bdC}{\mathbf{C}}

\newcommand{\bdH}{\mathbf{H}}
\newcommand{\bdI}{\mathbf{I}}

\newcommand{\bdK}{\mathbf{K}}

\newcommand{\bdM}{\mathbf{M}}
\newcommand{\bdN}{\mathbf{N}}

\newcommand{\bdQ}{\mathbf{Q}}
\newcommand{\bdR}{\mathbf{R}}
\newcommand{\bdS}{\mathbf{S}}
\newcommand{\bdT}{\mathbf{T}}
\newcommand{\bdU}{\mathbf{U}}
\newcommand{\bdV}{\mathbf{V}}

\newcommand{\bdX}{\mathbf{X}}

\newcommand{\bdzeta}{\boldsymbol{\zeta}}
\newcommand{\bdOmega}{\boldsymbol{\Omega}}
\newcommand{\bdomega}{\boldsymbol{\omega}}

\newcommand{\bdtheta}{\boldsymbol{\theta}}

\newcommand{\bdxi}{\boldsymbol{\xi}}
\newcommand{\bdSigma}{\boldsymbol{\Sigma}}

\newcommand{\bdPhi}{\boldsymbol{\Phi}}
\newcommand{\bdXi}{\boldsymbol{\Xi}}
\newcommand{\bdpsi}{\boldsymbol{\psi}}
\newcommand{\bdPsi}{\boldsymbol{\Psi}}
\newcommand{\bdLambda}{\boldsymbol{\Lambda}}
\newcommand{\bdlambda}{\boldsymbol{\lambda}}

\newcommand{\bdvphi}{\boldsymbol{\vphi}}

\newcommand{\stackeq}[1]{\stackrel{\text{(#1)}}{=}}

\newcommand{\stackpreceq}[1]{\stackrel{\text{(#1)}}{\preceq}}

\newcommand{\comma}{\text{,}}

\crefname{equation}{}{}
\crefname{figure}{Fig.}{Figs.}
\crefname{table}{TABLE}{TABLE}
\crefname{myassum}{Assumption}{Assumptions}
\crefname{myprop}{Proposition}{Propositions}
\crefname{mycorollary}{Corollary}{Corollarys}
\crefname{mylemma}{Lemma}{Lemmas}
\crefname{mytheorem}{Theorem}{Theorems}
\Crefname{secinapp}{Appendix}{Appendices}

\begin{document}

\title{Ergodic Sum Rate Capacity Achieving Transmit Design for Massive MIMO LEO Satellite Uplink Transmission}

\author{
Ke-Xin~Li,~\IEEEmembership{Member,~IEEE,} 
~Xiqi~Gao,~\IEEEmembership{Fellow,~IEEE,} 
and~Xiang-Gen~Xia,~\IEEEmembership{Fellow,~IEEE}
\thanks{Ke-Xin Li and Xiqi Gao are with the National Mobile Communications Research Laboratory (NCRL), Southeast University, Nanjing 210096, and are also with the Purple Mountain Laboratories (PML), Nanjing 211111, China (e-mail: likexin3488@seu.edu.cn, xqgao@seu.edu.cn).
Xiang-Gen Xia is with the Department of Electrical and Computer Engineering, University of Delaware,	Newark, DE 19716 USA (e-mail: xianggen@udel.edu). \emph{(Corresponding author: Xiqi Gao.)}}
}

\maketitle

\begin{abstract}
	In this paper, we investigate the ergodic sum rate (ESR) capacity achieving uplink (UL) transmit design for massive multiple-input multiple-output (MIMO) low-earth-orbit (LEO) satellite communications with statistical channel state information at the user terminals (UTs). The UL massive MIMO LEO satellite channel model with uniform planar array configurations at the satellite and UTs is presented. We prove that the rank of each UT's optimal transmit covariance matrix does not exceed that of its channel correlation matrix at the UT side, which reveals the maximum number of independent data streams transmitted from each UT to the satellite. We then prove that the transmit covariance matrix design can be transformed into the lower-dimensional matrix design without loss of optimality. We also obtain a necessary and sufficient condition when single data stream transmission from each UT to the satellite can achieve the ESR capacity. A conditional gradient (CG) method is developed to compute the ESR capacity achieving transmit covariance matrices. Furthermore, to avoid the exhaustive sample average, we utilize an asymptotic expression of the ESR and devise a simplified CG method to compute the transmit covariance matrices, which can approximate the ESR capacity. Simulations demonstrate the effectiveness of the proposed approaches.
\end{abstract}


\IEEEpeerreviewmaketitle


\section{Introduction} \label{section_introduction}
To serve vast areas with insufficient terrestrial network coverage, satellite communications (SATCOM) will be an indispensable part of next generation wireless network \cite{Oltjon2021SatelliteSpace}, and have been extensively investigated in the non-terrestrial networks (NTN) of 5G new radio (NR) \cite{3GPP_NonTerrestrial}.
In particular, low earth orbit (LEO) satellites, deployed between $200$ km to $2000$ km altitudes, have been recognized as a promising infrastructure to provide low-latency and ubiquitous broadband services for the user terminals (UTs) which have no access to the terrestrial network \cite{Di2019UltraDenseLEO,Xiao2022LEOSatellite}. Nowadays, commercial plans have been proposed to construct the mega-constellation by launching a huge number of LEO satellites into the space, e.g., Starlink \cite{Liu2021LEOSatellite}. LEO SATCOM has become a hotspot of research in both academia and industry. 

In the uplink (UL) SATCOM, multiple UTs on ground send messages to the satellite simultaneously.
The conventional UL multibeam SATCOM with single antenna configuration at the UT sides has been investigated in the literature, most of which focus on the capacity analysis, e.g., \cite{Christopoulos2011CapacityMultibeamUplink,Yang2016UpperBoundGEOMSCUplink,Arnau2014PerformanceReturnLink}.
The lower and upper bounds on the ergodic sum rate (ESR) capacity of the UL multibeam SATCOM were derived in \cite{Christopoulos2011CapacityMultibeamUplink} and \cite{Yang2016UpperBoundGEOMSCUplink}, respectively. In \cite{Arnau2014PerformanceReturnLink}, the ergodic capacity and outage capacity were analyzed by considering the spatially correlated rain attenuation in the UL multibeam satellite channels. In addition, the user scheduling combined with modulation and coding scheme selection for the UL of multibeam satellites with two-color reuse was investigated in \cite{Couble2018TwoColor}. The performance of UL multibeam SATCOM with multiple terrestrial relays was analyzed in \cite{Guo2019PerformanceUplink}, by taking account of the hardware impairments.

More flexible payloads have always been the pursuit of advanced SATCOM systems. In the last decades, massive multiple-input multiple-output (MIMO) has made great success in the terrestrial 5G system, and will continue to play an important role in the future 6G system \cite{CXWang2023OnRoad6G}.
Recently,  a massive MIMO transmission framework for LEO SATCOM was proposed in \cite{You2019MassiveMIMOLEO}, in which the channel model, multi-user precoder and detector, and the user grouping strategy were investigated. 
The multi-user precoder and detector in \cite{You2019MassiveMIMOLEO} are based on the  statistical channel state information (sCSI) instead of the  instantaneous CSI (iCSI).
This is because sCSI can remain stable for a relatively long time interval compared with the rapidly changing iCSI.
Since then, there have been several works focusing on the downlink (DL) transmission in massive MIMO LEO SATCOM, e.g., see \cite{Angeletti2020MassiveMIMOSatellite,KX2022DownlinkMassiveMIMOLEO,Roper2022Beamspace}.
The authors in \cite{Angeletti2020MassiveMIMOSatellite} compared the performance of different precoding techniques with the multibeam selection scheme, and devised a resource allocation approach to further improve the throughput.
In \cite{KX2022DownlinkMassiveMIMOLEO}, the single data stream transmission for each UT was shown to be optimal in the sense of maximizing the DL ESR, even though each UT is equipped with multiple antennas. The authors in \cite{Roper2022Beamspace} investigated the distributed linear precoder  and ground station (GS) equalizer for multi-satellite communications, which only rely on the positional information of the satellites and GS. 
In addition, a distributed massive MIMO was also introduced into LEO SATCOM, e.g., see \cite{Abdelsadek2022Distributed,Abdelsadek2023Broadband}, which can further enhance the system performance compared with the collocated counterparts.

Unlike the DL transmission, the massive LEO satellite UL receives much less attention. 
In \cite{You2019MassiveMIMOLEO}, an UL multi-user detector at the satellite side was designed with each UT equipped with a single antenna.
The authors in \cite{Shen2022RandomAccess} studied the joint channel estimation and device activity detection  for grant-free random access in massive MIMO LEO SATCOM, by using the orthogonal time-frequency space (OTFS) modulation to combat the large delays and Doppler shifts. 
In \cite{Zhou2023ActiveTerminal}, the active terminal identification, channel estimation and multi-user detection were jointly considered for grant-free non-orthogonal multiple access (NOMA) with OTFS modulation in massive MIMO LEO SATCOM systems.
However, to our best knowledge, the ESR capacity achieving UL transmit design for massive MIMO  LEO satellite systems with multiple antenna configurations at both the satellite and UTs has not been investigated.

In this paper, we study the UL transmit design that uses sCSI at the transmitter (sCSIT) in massive MIMO LEO SATCOM systems, where both the satellite and the UTs are equipped with uniform planar arrays (UPAs).
First, we present the UL massive MIMO LEO satellite channel model. The large propagation delays and Doppler shifts are pre-compensated at the UTs, to keep the received signals at the satellite synchronized, thereby supporting the orthogonal frequency division multiplexing (OFDM) based transmission.
 Then, we investigate the optimal UL transmit design for achieving the ESR capacity, by using the long-term sCSIT. 
We prove that the rank of each UT's transmit covariance matrix that achieves the UL ESR capacity is no larger than that of its own channel correlation matrix at the UT side, thus revealing the maximum number of independent data streams delivered from each multi-antenna UT to the satellite.
Moreover, we prove that each UT's transmit covariance matrix can be represented by a lower-dimensional matrix, so that the transmit covariance matrix design can be transformed into the lower-dimensional matrix design without any loss of optimality. 
We also obtain a necessary and sufficient condition when single data stream transmission from each UT to the satellite can achieve the ESR capacity.
We then develop a conditional gradient (CG) method to compute the ESR capacity achieving transmit covariance matrices. Furthermore, to avoid the complicated sample average, we utilize an asymptotic expression of the ESR, and devise a simplified CG method to compute the transmit covariance matrices, which can attain a near performance to the ESR capacity.

The remainder of this paper is organized as follows. \Cref{section_system_model} introduces the system model, where the UL channel model is presented for the satellite and the UTs both equipped with UPAs. \Cref{section_capacity_achieve_transmit_design} presents our main results on the UL transmit design, including the rank property of transmit covariance matrices, lower-dimensional matrix representation of transmit covariance matrices, and the CG methods to compute  the transmit covariance matrices. \Cref{section_simulation} provides the simulation results, and \Cref{section_conclusion} concludes this paper.

\textit{Notations:} Throughout this paper, lower case letters denote scalars, and boldface lower (upper) letters denote vectors (matrices). The set of all $n$-by-$m$ complex (real) matrices is denoted as $\bbC^{n\times m}$ ($\bbR^{n\times m}$). The trace, determinant, rank, conjugate, transpose, and conjugate transpose of a matrix are represented by $\trace(\cdot)$, $\det(\cdot)$, $\rank(\cdot)$, $(\cdot)^*$, $(\cdot)^T$, and $(\cdot)^H$, respectively. The Euclidean norm of vector $\bdx$ is denoted as $\norm{\bdx} = \sqrt{\bdx^H \bdx}$. The identity matrix is represented by $\bdI$ or $\bdI_n$. $\bdone$ and $\bdzro$ denote all-one and all-zero vectors, respectively. Denote $\otimes$ and $\odot$ as the Kronecker product and Hadamard product, respectively. Let $[\bdA]_{n,m}$ represent the $(n,m)$th element of matrix $\bdA$. $\Upsilon_{\max}(\bdA)$ denotes the maximum eigenvalue of $\bdA$. The diagonal matrix with $\bdx$ along its main diagonal is denoted as $\diag(\bdx)$. $\bbE \{ \cdot \}$ means mathematical expectation. $\clCN(\bdm,\bdC)$ denotes the proper complex Gaussian random vector with mean vector $\bdm$ and covariance matrix $\bdC$.
The uniform distribution on interval $[a, b]$ is denoted as $\mathrm{U} (a,b)$. 

\section{System Model} \label{section_system_model}

In this section, we first present the system configuration for UL massive MIMO LEO  SATCOM with OFDM modulation. Then, we derive the signal and channel models in the frequency domain of OFDM transmission after performing the Doppler and delay pre-compensation at the UT sides. The statistical properties of satellite channels are also provided.

\subsection{System Configuration}
\begin{figure}[!t]
	\centering
	\includegraphics[width=0.5\textwidth]{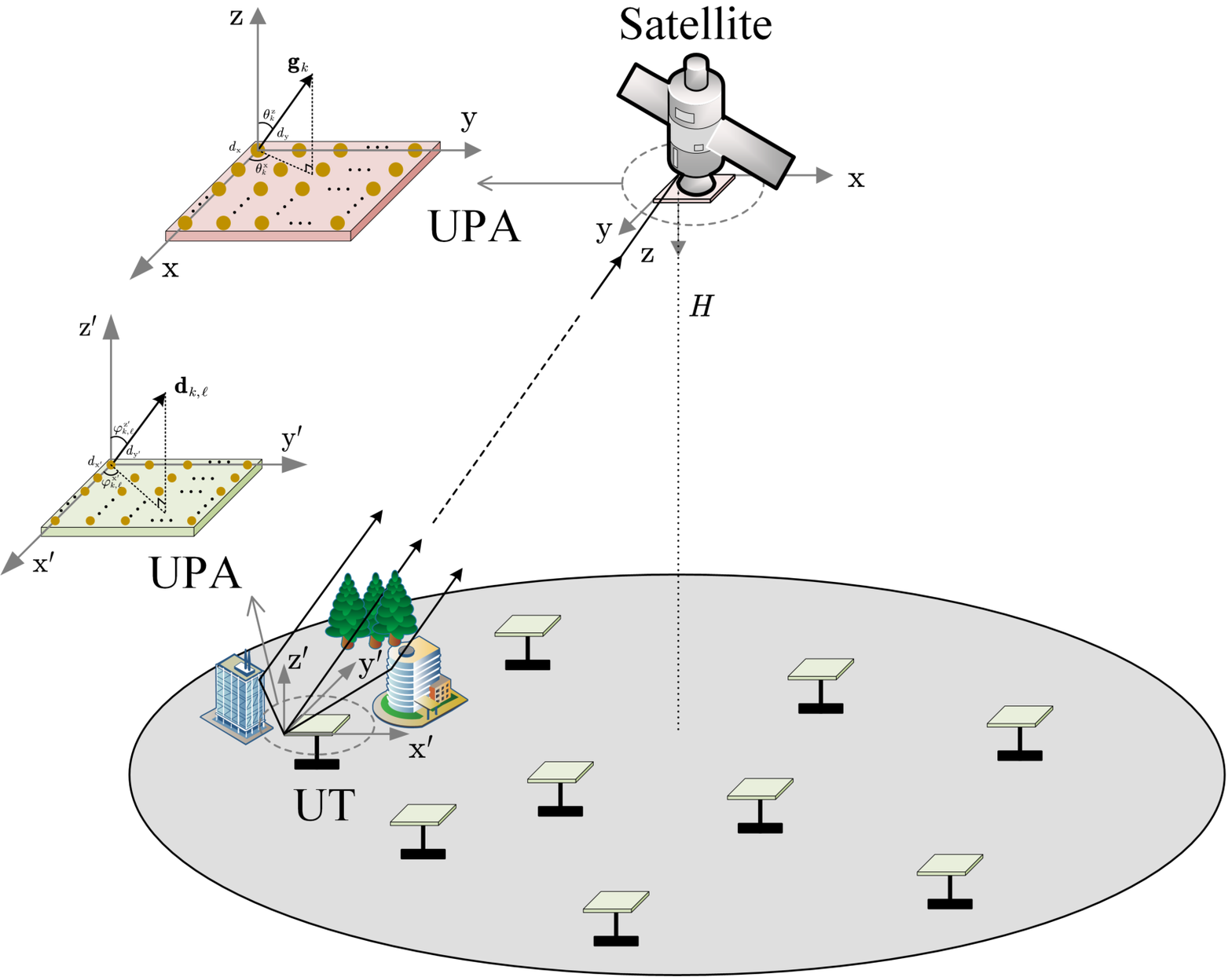}
	\caption{A massive MIMO LEO SATCOM system.}
	\label{fig_uplink_UPA}
\end{figure}

A massive MIMO LEO SATCOM system operating at the lower frequency bands, e.g., L/S/C bands, is considered. As depicted in \Cref{fig_uplink_UPA}, the mobile UTs on the ground send messages to the LEO satellite at an altitude of $H$.
The satellite and the mobile UTs are all equipped with UPAs. The UPA at the satellite has $\Mx$ and $\My$ directional elements in the $\rx$-axis and $\ry$-axis, respectively. Thus, the satellite has $\Mx \My \triangleq M$ antennas.
Meanwhile, each UT uses the UPA consisting of $\Nx$ and $\Ny$ omnidirectional antenna elements in the $\rx'$-axis and $\ry'$-axis, respectively. Hence, there are $\Nx \Ny \triangleq N$ antennas at each UT. 
Note that the antenna configurations can be extended to the case when the UPAs of the UTs have different numbers of antenna elements.

The OFDM modulation is used for wideband transmission in the LEO SATCOM system.
The number of subcarriers and cyclic prefix (CP) length are represented by $\Nsc$ and $\Ncp$, respectively. The subcarrier spacing is denoted by $\Delta f$, and then $\Ts = 1/(\Nsc \Delta f)$ is the system sampling period. The time duration of CP is given by $\Tcp = \Ncp \Ts$. The time intervals of one OFDM symbol excluding and including CP are given by $\Tsc = \Nsc \Ts$ and $T = \Tcp + \Tsc$, respectively.

\subsection{Signal and Channel Models} \label{subsection_UL_channel_model}
We consider the UL massive MIMO transmission. Let $\bdx_k(t) \in \Complex{N}{1}$ denote the transmit signal of UT $k$. The received signal at the satellite at time instant $t$ is given by
\begin{equation}
	\bdy(t) = \sum_{k} \int_{-\infty}^{\infty} \ckbdH_k(t,\tau) \bdx_k(t-\tau) \dint \tau + \bdz(t)\comma \label{RxSignal_time-domain}
\end{equation}
where $\ckbdH_k(t,\tau) \in \Complex{M}{N}$ is the channel impulse response of UT $k$, and $\bdz(t) \in \Complex{M}{1}$ is the additive noise at the satellite.
The time-varying channel impulse response $\ckbdH_k(t,\tau)$ can be written as
\begin{equation}
	\ckbdH_k(t,\tau) = \sum_{\ell=0}^{L_k-1} \check{a}_{k,\ell} e^{j2\pi f_{k,\ell} t } \delta(\tau-\tau_{k,\ell}) \bdg_{k,\ell} \bdd_{k,\ell}^H\comma \label{channel_model_UT_k}
\end{equation}
where $\delta(x)$ is the Dirac delta function, $j\triangleq \sqrt{-1}$, $L_k$ is the multipath number, $\check{a}_{k,\ell}$, $f_{k,\ell}$ and $\tau_{k,\ell}$ are the complex channel gain, the Doppler shift and the propagation delay for the $\ell$th path of UT $k$'s channel. In addition, $\bdg_{k,\ell} \in \Complex{M}{1}$ and $\bdd_{k,\ell} \in \Complex{N}{1}$ are the array response vectors for the $\ell$th path of UT $k$'s channel at the satellite and the UT sides, respectively. 

In LEO satellite channels, the Doppler shifts $f_{k,\ell}$'s and the propagation delays $\tau_{k,\ell}$'s are much larger than those in terrestrial wireless channels. 
Hence, the dominant parts in $f_{k,\ell}$'s and $\tau_{k,\ell}$'s need to be well pre-compensated at each UT, so that the received OFDM signals of the UTs can be synchronized at the satellite.
The Doppler shift $f_{k,\ell}$  can be separated as $f_{k,\ell} = f_{k,\ell}^{\sat} + f_{k,\ell}^{\ut}$, where $f_{k,\ell}^{\sat}$ and $f_{k,\ell}^{\ut}$ are the Doppler shifts caused by the motion of the satellite and UT $k$, respectively. In addition, the Doppler shifts $f_{k,\ell}^{\sat}$, $0\le \ell \le L_k-1$, tend to be identical for different paths of UT $k$'s channel \cite{3GPP_NonTerrestrial,Houcine2022NBIoTLEO,LiYou2022HybridLEO}. Thus, we can rewrite $f_{k,\ell}^{\sat} = f_{k}^{\sat}$, $0\le \ell  \le L_k-1$. Likewise, the  propagation delay $\tau_{k,\ell}$ can also be written as $\tau_{k,\ell} = \tau_{k}^{\sat} + \tau_{k,\ell}^{\ut} $, where $\tau_k^{\sat}$ is the large propagation delay due to the long distance between the satellite and UT $k$, and $\tau_{k,\ell}^{\ut}$ is the residual propagation delay depending on the scatter distribution around UT $k$. Notice that  $f_k^{\sat}$ and $\tau_k^{\sat}$ are the dominant parts in $f_{k,\ell}$ and $\tau_{k,\ell}$, respectively, which can be determined by the locations of the satellite and UT $k$.

Denote $\bdtheta_{k,\ell} = (\theta_{k,\ell}^{\rx},\theta_{k,\ell}^{\rz})$ and $\bdvphi_{k,\ell} = (\vphi_{k,\ell}^{\rx'},\vphi_{k,\ell}^{\rz'})$ as the paired angles-of-arrival (AoAs) and angles-of-departure (AoDs) related to the $\ell$th path of UT $k$'s channel. Then, $\bdg_{k,\ell}$ and $\bdd_{k,\ell}$ in \eqref{channel_model_UT_k} can be expressed as $\bdg_{k,\ell} = \bdg(\bdtheta_{k,\ell})$ and $\bdd_{k,\ell} = \bdd(\bdvphi_{k,\ell})$, respectively. Here, $\bdg(\bdtheta)$ and $\bdd(\bdvphi)$ for arbitrary $\bdtheta=(\theta_{\rx},\theta_{\rz})$ and $\bdvphi=(\vphi_{\rx'},\vphi_{\rz'})$ are defined as $\bdg(\bdtheta) = \bda_{\Mx} \left( \sin \theta_{\rz} \cos \theta_{\rx} \right) \otimes \bda_{\My} \left( \sin\theta_{\rz} \sin \theta_{\rx} \right)$ and $\bdd(\bdvphi) =  \bda_{\Nx} \left( \sin \vphi_{\rz'} \cos \vphi_{\rx'} \right) \otimes \bda_{\Ny} \left( \sin \vphi_{\rz'} \sin \vphi_{\rx'} \right)$, respectively.
Here, $\bda_{\nv} \left( x \right) = \frac{1}{\sqrt{\nv}} \left[ 1\ e^{-j \frac{2\pi d_{\rv} }{ \lambda } x }\ \cdots\  e^{-j\frac{2\pi d_{\rv} }{ \lambda } (\nv-1) x } \right]^T  \in \Complex{\nv}{1}$,
where $\lambda=c/f_c$ is the carrier wavelength, $c$ is the speed of the light, $f_c$ is the carrier frequency, $d_{\rv}$ is the spacing between adjacent antennas along the $\rv$-axis with $\rv \in \left\{ \rx,\ry,\rx',\ry' \right\}$.
Moreover, owing to the high altitude of the satellite, the paired AoAs for different paths of UT $k$'s channel are nearly identical, i.e., $\bdtheta_{k,\ell} = \bdtheta_k$, $0 \le \ell \le L_k-1$ \cite{3GPP_NonTerrestrial,ZhixiangGao2021SumRate}. 
In other words, the angular spread of each UT's channel observed at the satellite side is zero \cite{3GPP_NonTerrestrial,You2019MassiveMIMOLEO,KX2022DownlinkMassiveMIMOLEO}. 
Thus, we can discard the subscript of the path $\ell$ in $\bdg_{k,\ell}$ and rewrite it as $\bdg_{k,\ell} = \bdg_{k} = \bdg(\bdtheta_k)$,
where $\bdtheta_k = (\theta_k^{\rx},\theta_k^{\rz})$ is referred to as the physical AoA pair of UT $k$.
The AoAs and AoDs associated with the $\ell$th path of UT $k$'s satellite channel are illustrated in \Cref{fig_uplink_UPA}.
Notice that $\theta_k^{\rz}$ is also known as the nadir angle of UT $k$ \cite{Maral2020SatelliteCommunications}.
Furthermore, we define $\bdzeta_k = (\zeta_k^{\rx},\zeta_k^{\ry})$ as the space angle pair of UT $k$, where $\zeta_k^{\rx} = \sin \theta_k^{\rz} \cos \theta_k^{\rx} $ and $\zeta_k^{\ry} = \sin \theta_k^{\rz} \sin \theta_k^{\rx} $. 
Since the satellite is far away from the UTs, the space angle pairs $\{\bdzeta_k\}_{k=1}^K$ vary rather slowly and only depend on the locations of the satellite and the UTs. As soon as the UTs acquire their own location information, e.g., via the global navigation satellite system (GNSS), they can immediately derive the channel parameters $\{f_k^{\sat}, \tau_k^{\sat}, \bdzeta_k\}_{k=1}^K$ by utilizing the ephemeris.

Let $\{ \bdx_{k,s,r} \}_{r=0}^{\Nsc-1}$ denote the frequency-domain transmit signal of UT $k$ within the $s$th OFDM symbol. Then, the time-domain transmit signal is given by \cite{YongZuo2023OFDMBased}
\begin{align}
\bdx_{k,s}(t) = \sum_{r=0}^{\Nsc-1} \bdx_{k,s,r} e^{j2\pi r \Delta f \cdot t}\comma 
\end{align}
where $-\Tcp \le t-sT<\Tsc$.
Let $f_k^{\syn} = f_{k}^{\sat}$ and $\tau_k^{\syn} = \tau_{k}^{\min}$. By resorting to the time and frequency compensation techniques\cite{You2019MassiveMIMOLEO}, the transmit signal of UT $k$ in the $s$th OFDM symbol is given by 
\begin{align}
\bdx_{k,s}^{\syn}(t) = \bdx_{k,s} \left( t+\tau_k^{\syn} \right) e^{-j2\pi f_k^{\syn} \left( t + \tau_k^{\syn} \right) }.
\end{align}
The time-domain received signal at the satellite in the $s$th OFDM symbol can be written as
\begin{align}
	\bdy_{s}^{\syn}(t) &= \sum_{k} \int_{-\infty}^{\infty} \ckbdH_{k}(t,\tau) \bdx_{k,s}^{\syn}(t-\tau) \dint \tau + \bdz_{s}(t) \notag \\
&= \sum_{k} \bdg_k \sum_{\ell=0}^{L_k-1} \breve{a}_{k,\ell} e^{j2\pi f_{k,\ell}^{\ut} t} \bdd_{k,\ell}^H \cdot \bdx_{k,s}(t-\tau_{k,\ell}^{\ut})   + \bdz_{s}(t) \notag \\
	&= \sum_k \int_{-\infty}^{\infty} \breve{\bdH}_k(t,\tau) \bdx_{k,s}(t-\tau) \dint \tau + \bdz_{s}(t)\comma \label{rx_signal_compensate}
\end{align}
where $\breve{a}_{k,\ell} = \check{a}_{k,\ell} e^{j2\pi f_{k}^{\syn} \tau_{k,\ell}^{\ut}}$, and $\bdz_{s}(t)\in\Complex{M}{1}$ is the additive noise. In addition, $\breve{\bdH}_k(t,\tau)$ is the effective channel impulse response of UT $k$, and it can be written as
\begin{align}
	\breve{\bdH}_k(t,\tau) =  \bdg_{k} \cdot \sum_{\ell=0}^{L_k-1} \breve{a}_{k,\ell} e^{j2\pi f_{k,\ell}^{\ut} t } \delta(\tau-\tau_{k,\ell}^{\ut}) \bdd_{k,\ell}^H.
\end{align}
In contrast to the original channel impulse response $\ckbdH_k(t,\tau)$, the Doppler shifts and propagation delays in the effective channel impulse response $\breve{\bdH}_k(t,\tau)$ have been mitigated to a large extent in the sense that the time and frequency at the satellite and the UTs can be assumed as perfectly synchronized. Furthermore, the equivalent channel can be approximately treated as block-fading.

Let  $\bdH_k(t,f)$ denote the effective channel frequency response of UT $k$, which can be given by
\begin{align}
\bdH_k(t,f) &= \int_{-\infty}^{\infty} \breve{\bdH}_k(t,\tau) e^{-j2\pi f \tau} \dint \tau \notag \\
&= \bdg_k \left(\bdd_k(t,f)\right)^H\comma
\end{align}
where $\bdd_k(t,f) = \sum_{\ell=0}^{L_k-1} \breve{a}_{k,\ell}^* e^{ -j2\pi ( f_{k,\ell}^{\ut} t - f \tau_{k,\ell}^{\ut} ) } \bdd_{k,\ell} \in \Complex{N}{1}$.
Thus, the frequency-domain received signal at the satellite on the $r$th subcarrier in the $s$th OFDM symbol can be written as  \cite{YongZuo2023OFDMBased}
\begin{align}
	\bdy_{s,r} &= \xinv{\Tsc} \int_{sT}^{sT+\Tsc} \bdy_{s}^{\syn}(t) e^{-j2\pi r \Delta f \cdot t } \dint t \notag \\
	&= \sum_{k} \bdH_{k,s,r} \bdx_{k,s,r} + \bdz_{s,r} \comma \label{RxSignal_ysr_Hksr_xksr}
\end{align}
where $\bdH_{k,s,r}$ is the channel matrix of UT $k$, $\bdz_{s,r}$ is the additive Gaussian noise, both on the $r$th subcarrier of the $s$th OFDM symbol. Furthermore, $\bdH_{k,s,r}$ in \eqref{RxSignal_ysr_Hksr_xksr} can be written as
\begin{equation}
	\bdH_{k,s,r} = \bdH_k(sT,r\Delta f) = \bdg_k \bdd_{k,s,r}^H\comma
\end{equation}
where $\bdd_{k,s,r} = \bdd_k(sT,r\Delta f)$.


\subsection{Statistical Properties of Satellite Channels}
In this subsection, we briefly describe the statistical properties of the massive MIMO LEO satellite UL channels. For convenience, we omit the subscripts of OFDM symbol $s$ and subcarrier $r$ in $\bdH_{k,s,r} = \bdg_k \bdd_{k,s,r}^H$ and denote $\bdH_k = \bdg_k \bdd_k^H$ as the flat fading channel matrix of UT $k$ on a specific subcarrier over an OFDM symbol. 
Based on the effective physical channel models in \Cref{subsection_UL_channel_model}, $\bdg_k$ is a non-random vector, and $\bdd_k$ can be modeled according to the Rician distribution as 
\begin{align}
	\bdd_k = \sqrt{\frac{\kappa_k \beta_k}{\kappa_k+1}} \bdd_{k,0} + \sqrt{\frac{\beta_k}{\kappa_k + 1}} \tdbdd_k\comma \label{Rician_fading_dk}
\end{align}
where $\kappa_k$ is the Rician factor, $\beta_k = \bbE \left\{ \trace(\bdH_k \bdH_k^H) \right\} = \bbE \left\{ \snorm{\bdd_k}^2 \right\}$ is the average channel power, $\bdd_{k,0}$ represents the direction of the line-of-sight (LoS) path seen at UT $k$'s side, and $\tdbdd_k$ is distributed as $\tdbdd_k \sim \clCN(\bdzro, \bdSigma_k)$ with $\trace(\bdSigma_k) = 1$. In addition, we assume that the random vectors $\bdd_k$'s are independent for different $k$'s.  The channel matrix $\bdH_k$ can be written as
\begin{align}
\bdH_k = \bdg_k \bdd_k^H = \sqrt{\frac{\kappa_k \beta_k}{\kappa_k+1}} \bdH_k^{\LoS} + \sqrt{\frac{\beta_k}{\kappa_k + 1}} \bdH_k^{\NLoS}\comma
\end{align}
where $\bdH_k^{\LoS} = \bdg_k \bdd_{k,0}^H$ is the deterministic LoS component, $\bdH_k^{\NLoS} = \bdg_k \tdbdd_k^H$ is the random scattering component. 
The channel correlation matrices of UT $k$ at the satellite and the UT sides are given by 
\begin{align}
\bdR_k^{\sat} &= \bbE \left\{ \bdH_k \bdH_k^H \right\} = \beta_k \bdg_k \bdg_k^H \\
\bdR_k^{\ut}  &= \bbE \left\{ \bdH_k^H \bdH_k \right\} \notag \\
 &= \bbE \left\{ \bdd_k \bdd_k^H \right\} 
 =  \frac{\kappa_k \beta_k}{\kappa_k+1} \bdd_{k,0} \bdd_{k,0}^H + \frac{\beta_k}{\kappa_k + 1} \bdSigma_k\comma \label{Rkut_CCM}
\end{align}
respectively.
It is worth noting that $\bdR_k^{\sat}$ is rank-one indicating that the arriving signals on different antennas at the satellite are highly correlated. Meanwhile, the rank of $\bdR_k^{\ut}$ is determined by the scatter distribution around UT $k$.

\section{UL Transmit Design} \label{section_capacity_achieve_transmit_design}

In this section, with the above established massive MIMO LEO satellite channel model, we investigate the transmit covariance matrix design that achieves the UL ESR capacity. First, we prove that the rank of each UT's transmit covariance matrix should be no larger than that of its channel correlation matrix at the UT side. Then, we prove that each UT's transmit covariance matrix can be represented by a low-dimensional matrix, so that the transmit covariance matrix design can be transformed into the lower-dimensional matrix design without any loss of optimality. We further derive a necessary and sufficient condition when single data stream transmission from each UT to the satellite can achieve the ESR capacity. Afterwards, a CG method is developed to compute the ESR capacity achieving transmit covariance matrices with guaranteed convergence. Further, in order to avoid the complicated sample average, we resort to an asymptotic expression of the ESR, and then devise a simplified CG method to compute the transmit covariance matrices, which can attain a near performance to the ESR capacity.

\subsection{Rank Property of Transmit Covariance Matrices} \label{subsec_rank_property}
For convenience, we omit the subscripts of OFDM symbol $s$ and subcarrier $r$ in $\bdx_{k,s,r}$, and denote $\bdx_k \in \Complex{N}{1}$ as the transmit signal of UT $k$ on a specific subcarrier over an OFDM symbol.
In this paper, we consider a general design of the transmit signals $\bdx_k$'s, where $\bdx_k$ is a circularly symmetric complex Gaussian (CSCG) random vector with zero mean and covariance matrix $\bdQ_k = \bbE \{ \bdx_k \bdx_k^H \} \in \Complex{N}{N}$. Let us denote the eigenvalue decomposition (EVD) of $\bdQ_k$ as $\bdQ_k = \bdV_{\rQ,k} \bdLambda_{\rQ,k} \bdV_{\rQ,k}^H$, where the column vectors in $\bdV_{\rQ,k} \in \Complex{N}{N}$ are the eigenvectors and $\bdLambda_{\rQ,k} \in \Real{N}{N}$ is a diagonal matrix with the non-negative eigenvalues as the diagonal elements. Then, the transmit signal $\bdx_k$ can be written as
\begin{align}
	\bdx_k = \bdV_{\rQ,k} \bdLambda_{\rQ,k}^{1/2} \bds_{\rQ,k}\comma
\end{align}
where $\bds_{\rQ,k} \sim \clCN(\bdzro,\bdI_{N})$ denotes the data vector of UT $k$.
Moreover, we also consider the sum power constraint $\trace (\bdQ_k) \le P_k$ for each UT $ k$ in the UL transmission.
We assume that $K$ mobile UTs send messages to the satellite simultaneously. The UT index set is denoted by $\clK = \left\{1,\dots,K\right\}$.
Thus, the received signal $\bdy \in \Complex{M}{1}$ at the satellite is expressed as
\begin{equation}
	\bdy = \sum_{k=1}^K \bdH_k \bdx_k  + \bdz\comma \label{Data_Transmission_Model_UL}
\end{equation}
where $\bdz \in \Complex{M}{1}$ is the additive Gaussian noise at the satellite  distributed as $\bdz \sim \clCN (\bdzro, \sigma^2 \bdI_M)$.

We assume that perfect iCSI is known by the receiver at the satellite side, while only sCSI is known by the transmitters at the UTs' side.
The UL ESR capacity $C$ is given by
\begin{align}
	C ={}& \max_{\bdQ_k \succeq \bdzro\comma\ \trace(\bdQ_k) \le P_k \comma\ \forall k} \notag \\
	 &\quad \bbE \left\{ \log \det \left( \bdI_M + \xinv{\sigma^2} \sum_{k=1}^{K} \bdH_k \bdQ_k \bdH_k^H \right) \right\}. \label{UL_Capacity}
\end{align}
By substituting $\bdH_k = \bdg_k \bdd_k^H$ into \eqref{UL_Capacity}, $C$ can be further written as
\begin{align}
	C ={}& \max_{\bdQ_k \succeq \bdzro\comma\ \trace(\bdQ_k) \le P_k\comma\ \forall k} \notag \\
	&\quad \bbE \left\{ \log \det \left( \bdI_M + \xinv{\sigma^2} \sum_{k=1}^K \bdd_k^H \bdQ_k \bdd_k \cdot \bdg_k \bdg_k^H \right) \right\}. \label{UL_Capacity_dQd}
\end{align}

Although the problem in \eqref{UL_Capacity_dQd} is a convex optimization problem \cite{BoydConvexOptimization}, the mathematical expectation in the ESR makes it challenging to manifest the solution of the problem in \eqref{UL_Capacity_dQd}.

In fact, the rank of the matrix $\bdQ_k$ is the number of independent data streams that are delivered through UT $k$'s channel. In the following theorem, we show the rank property of the optimal transmit covariance matrices $\{\bdQ_k\}_{k=1}^K$, which paves the way for the lower-dimensional representation of $\{\bdQ_k\}_{k=1}^K$.
\begin{mytheorem} \label{Proposition_1}
	The transmit covariance matrices $\{\bdQ_k\}_{k=1}^K$ that achieve the UL ESR capacity should satisfy
	\begin{equation}
		\rank(\bdQ_k) \le \rank \left( \bdR_k^{\ut} \right)\comma \ \forall k \in \clK. \label{Rank_Qk_UL_Edd}
	\end{equation}
\end{mytheorem}
\begin{IEEEproof}
	Please refer to \Cref{appendix_Proposition_1_Proof}.
\end{IEEEproof}

The rank property of $\bdQ_k$ for the $k$th UT in \Cref{Rank_Qk_UL_Edd} holds independently of other UTs' channel correlation matrices. From \Cref{Proposition_1}, the maximum number of independent data streams transmitted from UT $k$ to the satellite should be no larger than the rank of UT $k$'s channel correlation matrix $\bdR_k^{\ut}$. It can be anticipated that if there are only sparse scatterers distributed around UT $k$, $\rank(\bdQ_k)$ may be much less than the number of antennas $N$ at UT $k$. 
Moreover, because the assumption on each UT's channel distribution is not invoked in the proof procedure of \Cref{Proposition_1}, the rank property in \Cref{Rank_Qk_UL_Edd} is applicable to a large set of channel distributions as long as each UT's channel is independently distributed.

Next, we show that in some extreme cases including the low signal-to-noise ratio (SNR) case and the high Rician factor case, the optimal $\{\bdQ_k\}_{k=1}^K$ to the problem  in \eqref{UL_Capacity_dQd} are of rank-one.

\subsubsection{Low SNR Case} \label{subsubsection_low_SNR_Qmat}
If $P_k \rightarrow 0$, $\forall k \in \clK$, holds, the UL ESR is reduced into $\xinv{\sigma^2} \sum_{k=1}^K \trace \left( \bdR_k^{\ut} \bdQ_k \right)$.
Then, the problem  in \eqref{UL_Capacity_dQd} can be simplified into
\begin{equation}
	C^{\LowSNR} = \max_{ \bdQ_k \succeq \bdzro\comma\ \trace(\bdQ_k) \le P_k\comma\ \forall k \in \clK }\ \sum_{k=1}^{K} \trace\left( \bdR_k^{\ut} \bdQ_k \right). \label{Problem_UL_sum_rate_capacity_lowSNR}
\end{equation}
The optimal matrices $\{\bdQ_k\}_{k=1}^K$ to the problem  in \Cref{Problem_UL_sum_rate_capacity_lowSNR} are given by
\begin{equation}
	\bdQ_k = P_k \cdot \bdb_{\rQ,k} \bdb_{\rQ,k}^H\comma\ \forall k \in \clK\comma \label{Qk_optimal_low_SNR}
\end{equation}
where $\bdb_{\rQ,k} \in \Complex{N}{1}$ is the unit-norm eigenvector of $\bdR_k^{\ut}$ associated with its maximum eigenvalue.

\subsubsection{High Rician  Factor Case} \label{subsubsection_high_Rician_factor_Qmat}
If $\kappa_k \rightarrow \infty$ holds for each UT $ k \in \clK$, the problem in \eqref{UL_Capacity_dQd} is reduced into
\begin{align}\label{Problem_UL_sum_rate_capacity_infRician}
C^{\HighRiceK} ={}& \max_{\bdQ_k \succeq \bdzro\comma\ \trace(\bdQ_k) \le P_k\comma\ \forall k \in \clK}\notag \\
&\quad  \log \det \left( \bdI_M + \xinv{\sigma^2} \sum_{k=1}^{K} \beta_k \bdd_{k,0}^H \bdQ_k \bdd_{k,0} \cdot \bdg_k \bdg_k^H \right).
\end{align}
The optimal matrices $\{\bdQ_k\}_{k=1}^K$ to the problem in \Cref{Problem_UL_sum_rate_capacity_infRician} can be derived as follows
\begin{equation}
	\bdQ_k = P_k \cdot \bdd_{k,0} \bdd_{k,0}^H,\ \forall k \in \clK. \label{Qk_optimal_high_Rician}
\end{equation}

The optimal transmit strategy for the high Rician factor case is to perform the transmit beamforming along each UT's LoS direction seen at the UT side.
In this case, owing to $\bdd_{k,0} = \bdd(\bdvphi_{k,0})$, only the paired AoDs $\bdvphi_{k,0}=(\vphi_{k,0}^{\rx'},\vphi_{k,0}^{\rz'})$ for the LoS path is required to be known at UT $k$.
Moreover, the relatively simple phased array antennas (PAAs) can be used at the UT sides to implement the beamformers $\{\sqrt{P_k} \bdd_{k,0}\}_{k=1}^K$, which can significantly reduce the implementation cost and complexity.

\subsection{Lower-Dimensional Matrix Representation of Transmit Covariance Matrix} \label{subsection_structure_transmit_matrix}
In this subsection, we show that each UT $k$'s transmit covariance matrix can be represented by a lower-dimensional matrix, and the transmit covariance matrix design can be transformed into the lower-dimensional matrix design.
Let us denote the EVD of $\bdSigma_k$ in \eqref{Rkut_CCM} as $\bdSigma_k = \bdU_k \diag(\bdlambda_k) \bdU_k^H$. The columns of $\bdU_k = [ \bdu_{k,1}\ \cdots \ \bdu_{k,S_k} ] \in \Complex{N}{S_k}$ are the eigenvectors and the elements of $\bdlambda_k =[ \lambda_{k,1},\dots,\lambda_{k,S_k}]^T$ are the corresponding positive eigenvalues in non-increasing order, where $S_k = \rank(\bdSigma_k)$.
Let us further denote the linear subspace spanned by the columns in $\bdU_k$ as $\spann(\bdU_k)$.
We can separate $\bdd_{k,0}$ into two orthogonal terms as
\begin{equation}
	\bdd_{k,0} = \udbdu_{k,0}  + \bdU_k \bdxi_{k,0}  \comma \label{dk0_uk0_pk0}
\end{equation}
where $\udbdu_{k,0} \triangleq (\bdI - \bdU_k \bdU_k^H) \bdd_{k,0}$ and $\bdxi_{k,0} \triangleq \bdU_k^H \bdd_{k,0}$. The first term in \eqref{dk0_uk0_pk0} is orthogonal to the linear subspace $\spann(\bdU_k)$ and the second term lies in $\spann(\bdU_k)$.
By using the Rician fading channel assumption in \eqref{Rician_fading_dk}, we can rewrite $\tdbdd_k$ as $\tdbdd_k = \bdU_k \udtdbdc_k$, where the elements in $\udtdbdc_k = \left[ \tdc_{k,1},\dots,\tdc_{k,S_k} \right]^T \in \Complex{S_k}{1}$ are independent CSCG random variables with distinct variances. Indeed, $\udtdbdc_k$ is distributed as $\udtdbdc_k \sim \clCN\left(\bdzro,\diag(\bdlambda_k)\right)$.
Henceforth, $\bdd_k$ can be rewritten as
\begin{align}
	\bdd_k & = \sqrt{\frac{\kappa_k \beta_k}{\kappa_k+1}} \bdd_{k,0} + \sqrt{\frac{\beta_k}{\kappa_k + 1}} \bdU_k \udtdbdc_k \notag                                                                  \\
	       & = \sqrt{\frac{\kappa_k \beta_k}{\kappa_k+1}} \udbdu_{k,0} + \sqrt{\frac{\kappa_k \beta_k}{\kappa_k+1}} \bdU_k \bdxi_{k,0} + \sqrt{\frac{\beta_k}{\kappa_k + 1}} \bdU_k \udtdbdc_k \notag \\
	       & = \sqrt{\frac{\beta_k}{\kappa_k + 1}} \left( \sqrt{\kappa_k} \udbdu_{k,0} + \bdU_k \left( \sqrt{\kappa_k} \bdxi_{k,0} + \udtdbdc_k \right) \right). \label{dk_Bkck}
\end{align}

Let $\eta_{k,0} = \snorm{\udbdu_{k,0}}^2$ and $\bdu_{k,0} = \frac{\udbdu_{k,0}}{\snorm{\udbdu_{k,0}}}$. Define $\bdB_k \in \Complex{N}{\Sone_k}$ and $\bdc_k \in \Complex{\Sone_k}{1}$ as follows
\begin{subequations} \label{Bk_ck}
	\begin{align}
		\bdB_k & =
		\begin{cases}
			\left[ \bdu_{k,0} \  \bdU_k \right]\comma & \text{ if } \udbdu_{k,0} \ne \bdzro\comma \\
			\bdU_k\comma                              & \text{ if } \udbdu_{k,0} = \bdzro\comma
		\end{cases} \\
		\bdc_k & =
		\begin{cases}
			\sqrt{\frac{\beta_k}{\kappa_k + 1}}\left[
			\begin{matrix}
					\sqrt{\kappa_k \eta_{k,0}} \\
					\sqrt{\kappa_k} \bdxi_{k,0} + \udtdbdc_{k}
				\end{matrix} \right] \comma                     & \text{ if } \udbdu_{k,0} \ne \bdzro\comma \\
			\sqrt{\frac{\beta_k}{\kappa_k + 1}} \left(\sqrt{\kappa_k} \bdxi_{k,0} + \udtdbdc_{k}\right) \comma & \text{ if } \udbdu_{k,0} = \bdzro\comma
		\end{cases}
	\end{align}
\end{subequations}
respectively, where $\Sone_k$ is defined as
\begin{align}
	\Sone_k = \begin{cases}
		 S_k + 1\comma & \text{ if } \udbdu_{k,0} \ne \bdzro\comma \\
		  S_k\comma & \text{ if } \udbdu_{k,0} = \bdzro.
	\end{cases}
\end{align}
From \Cref{dk_Bkck,Bk_ck}, we can rewrite $\bdd_k$ as
\begin{equation}
	\bdd_k = \bdB_k \bdc_k\comma \label{dk_Bk_ck_equality}
\end{equation}
where the columns in $\bdB_k$ are orthogonal to each other, i.e., $\bdB_k^H \bdB_k = \bdI$. 
Notice that $\bdc_k$ can be written as $\bdc_k = \sqrt{\frac{\kappa_k \beta_k}{\kappa_k+1}} \bdc_{k,0} + \sqrt{\frac{\beta_k}{\kappa_k + 1}} \tdbdc_k$, where $\bdc_{k,0}$ and $\tdbdc_k$ are given by
\begin{align}
	\bdc_{k,0} &=\begin{cases}
		[ \sqrt{\eta_{k,0}} \  \bdxi_{k,0}^T ]^T\comma & \text{ if } \udbdu_{k,0} \ne \bdzro\comma \\
		\bdxi_{k,0}\comma & \text{ if } \udbdu_{k,0} \ne \bdzro\comma
	\end{cases} \\
	\tdbdc_k &= \begin{cases}
		[0 \ \udtdbdc_k^T]^T\comma & \text{ if } \udbdu_{k,0} \ne \bdzro\comma \\
		\udtdbdc_k\comma & \text{ if } \udbdu_{k,0} = \bdzro\comma
	\end{cases}
\end{align}
respectively.
Moreover, $\bdR_k^{\ut}$ can be rewritten as
\begin{equation}
	\bdR_k^{\ut} = \bdB_k \bdOmega_k \bdB_k^H\comma \label{Rkut_Bk_Omegak_Bk}
\end{equation}
where $\bdOmega_k \triangleq \bbE\{\bdc_k \bdc_k^H\} \in \Complex{\Sone_k}{\Sone_k}$ is given by
\begin{equation}
	\bdOmega_k =
	\begin{cases}
		\frac{\beta_k}{\kappa_k + 1}\left[ \begin{matrix}
				\kappa_k \eta_{k,0}  \comma                 & \kappa_k \sqrt{\eta_{k,0}} \bdxi_{k,0}^H                \\
				\kappa_k \sqrt{\eta_{k,0}} \bdxi_{k,0} \comma & \tilde{\bdOmega}_k
			\end{matrix}\right]  \comma                             & \text{if } \udbdu_{k,0} \ne \bdzro\comma \\
		 \frac{\beta_k}{\kappa_k + 1} \tilde{\bdOmega}_k \comma & \text{if } \udbdu_{k,0} = \bdzro\comma
	\end{cases}
\end{equation}
where $\tilde{\bdOmega}_k = \kappa_k \bdxi_{k,0} \bdxi_{k,0}^H + \diag(\bdlambda_k)$.
Note that $\bdOmega_k$ is positive definite. Thus, we have
\begin{equation}
	\rank\left( \bdR_k^{\ut} \right) = \rank(\bdB_k \bdOmega_k \bdB_k^H) \stackeq{a} \rank(\bdB_k) \stackeq{b} \Sone_k \comma \label{rank_relation_Sk_Edd}
\end{equation}
where (a) comes from \cite[Observation 7.1.8(b)]{Horn2013MatrixAnalysis}, and (b) follows from the fact that $\bdB_k$ has orthogonal columns.

\begin{mytheorem} \label{Proposition_2}
	The optimal solution $\{\bdQ_k^{\star}\}_{k=1}^K$ to the problem in \eqref{UL_Capacity_dQd} can be obtained by
	\begin{equation} \label{Relation_Qk_Tk_optimal}
		\bdQ_k^{\star} =
		\bdB_k \bdT_k^{\star} \bdB_k^H\comma\ \forall k \in \clK.
	\end{equation}
	Here, $\{\bdT_k^{\star}\}_{k=1}^K$ is the optimal solution to the following problem
	\begin{align} \label{Problem_UL_sum_rate_Tk}
	C ={}& \max_{\bdT_k \succeq \bdzro\comma\ \trace(\bdT_k) \le P_k\comma \ \forall k}\notag \\
	&\quad \bbE \left\{ \log \det \left( \bdI_M + \xinv{\sigma^2} \sum_{k=1}^K \bdc_k^H \bdT_k \bdc_k \cdot \bdg_k \bdg_k^H \right) \right\}.
	\end{align}
\end{mytheorem}
\begin{IEEEproof}
	Please refer to \Cref{appendix_Proposition_2_proof}.
\end{IEEEproof}

\Cref{Proposition_2} reveals that each UT $k$'s $N \times N$ transmit covariance matrix $\bdQ_k$ can be represented by an $\Sone_k \times \Sone_k$ lower-dimensional matrix $\bdT_k$, whose dimension is exactly equal to $\rank(\bdR_k^\ut)$ as shown by \Cref{rank_relation_Sk_Edd}. Interestingly, this is consistent with the results in \Cref{Proposition_1}. 
It is worth noting that the proof of \Cref{Proposition_2} relies on the equality $\bdd_k = \bdB_k \bdc_k$ in \eqref{dk_Bk_ck_equality}, which is derived based on the Rician fading channel assumption in \eqref{Rician_fading_dk}.
With the aid of \Cref{Proposition_2}, the transmit covariance matrix design can be transformed into the lower-dimensional matrix design without any loss of optimality.
After the optimal lower-dimensional matrices $\{\bdT_k^{\star}\}_{k=1}^K$ are obtained, the optimal transmit covariance matrices $\{\bdQ_k^{\star}\}_{k=1}^K$ can be obtained immediately by using \eqref{Relation_Qk_Tk_optimal}. Henceforth, we only need to concentrate on the optimization of the lower-dimensional matrices $\{\bdT_k\}_{k=1}^K$.
The problem in \Cref{Problem_UL_sum_rate_Tk} keeps the convex property, and the optimization variables therein have lower dimensions. 

Let $\bdT_k = \bdV_{\rT,k} \bdLambda_{\rT,k} \bdV_{\rT,k}^H$ denote the EVD of $\bdT_k$, where the column vectors of $\bdV_{\rT,k} \in \Complex{\Sone_k}{\tdS_{k}}$ consist of eigenvectors and  $\bdLambda_{\rT,k} \in \Real{\tdS_k}{\tdS_k}$ is a diagonal matrix including the corresponding non-negative eigenvalues along the main diagonal. The transmit signal $\bdx_k$ should be given by 
\begin{align}
	\bdx_k = \bdB_k \bdV_{\rT,k} \bdLambda_{\rT,k}^{1/2} \bds_{\rT,k}\comma
\end{align}
where $\bds_{\rT,k} \sim \clCN(\bdzro,\bdI_{\tdS_k})$ denotes the lower-dimensional data vector of UT $k$.
Notice that the analysis of two extreme cases including the low SNR case and the high Rician factor case in \Cref{subsec_rank_property} after \Cref{Proposition_1} also holds here.

The following results show that the solution to the problem in \Cref{Problem_UL_sum_rate_Tk} can be further simplified under some special conditions.
\begin{mytheorem} \label{Proposition_3}
	If $\bdxi_{k,0} = \bdzro$ holds for UT $k$, then $\bdT_k^{\star}$ is a diagonal matrix.
\end{mytheorem}
\begin{IEEEproof}
	Please refer to \Cref{appendix_Prop_Tk_diagonal_proof}.
\end{IEEEproof}

In fact, $\bdxi_{k,0} = \bdzro$ indicates that $\bdd_{k,0}$ is orthogonal to $\spann(\bdU_k)$. In this case, $\bdB_k$ reduces to $\bdB_k = \left[\bdd_{k,0}\ \bdU_k\right]$, whose column vectors actually become the eigenvectors of $\bdQ_k^{\star}$. 
As shown in \Cref{Proposition_3}, if $\bdxi_{k,0} = \bdzro$ holds, the optimal transmit strategy of UT $k$ would be sending independent data streams along the directions determined by the columns in $\bdB_k = \left[\bdd_{k,0}\ \bdU_k\right]$.

In the following, we provide a necessary and sufficient condition when the rank-one transmit covariance matrix used by a specific UT can achieve the ESR capacity.

\begin{mytheorem} \label{Proposition_4}
	The optimal matrix $\bdT_k$ of the problem in \eqref{Problem_UL_sum_rate_Tk} is given by 
	$\bdT_k^{\star} = P_k \bdw_{k} \bdw_{k}^H$, if and only if 
	\begin{align}
		 &\Upsilon_{\max} \left( \bbE\left\{  \frac{  G_k}{ 1 + G_k P_k \abs{ \bdc_k^H \bdw_{k} }^2 } \bdc_k \bdc_k^H \right\} \right) \notag \\
		 ={}& \bbE\left\{  \frac{  G_k \sabs{\bdc_k^H \bdw_{k}}^2 }{ 1 + G_k P_k \abs{ \bdc_k^H \bdw_{k} }^2 } \right\} \comma \label{eigmax_wtk}
	\end{align}
with $G_k = \xinv{\sigma^2} \bdg_k^H \tdbdA_k^{-1} \bdg_k$, and $\tdbdA_k = \bdI_M + \xinv{\sigma^2} \sum_{i\ne k} \bdc_i^H \bdT_i \bdc_i \cdot \bdg_i \bdg_i^H$,  where $\bdw_{k}$ is a unit-norm vector.
\end{mytheorem}
\begin{IEEEproof}
	Please refer to \Cref{appendix_Prop_Tk_rankone_proof}.
\end{IEEEproof}

According to \Cref{Proposition_2}, if $\bdT_k^{\star} = P_k \bdw_{k} \bdw_{k}^H$, then $\bdQ_k^{\star} = \bdB_k \bdT_k^{\star} \bdB_k^H = P_k \bdB_k \bdw_{k} \bdw_{k}^H \bdB_k^H$.
The unit-norm vector $\bdB_k \bdw_{k}$ actually plays the role of the beamforming vector of UT $k$.
\Cref{Proposition_4} provides a necessary and sufficient condition when single data stream transmission from each UT to the satellite can achieve the UL ESR capacity. In other words, the optimal matrix $\bdT_k$ of the problem in \eqref{Problem_UL_sum_rate_Tk} has the rank-one structure if and only if there exists some vector $\bdw_k$ which fulfills the condition in \eqref{eigmax_wtk}.
With the help of \Cref{Proposition_2}, as long as  the condition in \eqref{eigmax_wtk} is satisfied, the lower-dimensional matrix design can be further simplified into the rank-one matrix design, i.e., vector design, for UT $k$, which can significantly reduce the implementation complexity especially at the UT sides. 

On the other hand, if there does not exist such a vector $\bdw_k$ that satisfies the condition in \eqref{eigmax_wtk}, the single data stream transmission from UT $k$ to the satellite will be unable to achieve the UL ESR capacity. Hence, it still needs to consider the lower-dimensional matrix design for the most general circumstances. In the next subsection, we elaborate the general design approaches for the low-dimensional matrices, which can achieve or approximate the UL ESR capacity for massive MIMO LEO SATCOM systems.

\subsection{CG Method for UL Transmit Design} \label{subsection_General_Design}
In this subsection, we develop the CG based iterative method, a.k.a., the Frank-Wolfe method \cite{Bertsekas2015ConvexAlgorithm}, to solve the convex optimization problem in \eqref{Problem_UL_sum_rate_Tk} with guaranteed convergence. Specifically, in each iteration, a simple linear optimization subproblem needs to be solved to obtain a feasible direction, and then the optimization variables can be updated along their feasible directions. Simulation results in \Cref{section_simulation} show that the CG method can converge to the optimal points of the problem in \eqref{Problem_UL_sum_rate_Tk} within very few iterations.

The gradient of the ESR $R_{\SUM}= \bbE \{ \log \det ( \bdI_M + \xinv{\sigma^2} \sum_{k=1}^K \bdc_k^H \bdT_k \bdc_k \cdot \bdg_k \bdg_k^H ) \}$ with respect to $\bdT_k$ can be written as follows
\begin{align}
	\bdM_k = \nabla_{\bdT_k} R_{\SUM}
	& = \bbE \left\{ \tilde{W}_k \bdc_k \bdc_k^H \right\} \comma \label{Rsum_gradient_Tk}
\end{align}
where $\tilde{W}_k = \frac{ G_k }{ 1 + G_k \bdc_k^H \bdT_k \bdc_k }$.
We use $(\cdot)^{(n)}$ to denote the argument in the $n$th iteration.
Given $\{\bdT_k^{(n)}\}_{k=1}^K$, a feasible solution to the problem in \eqref{Problem_UL_sum_rate_Tk} can be derived by solving the following linear programming problem \cite{Bertsekas2015ConvexAlgorithm}
\begin{align}
	\max_{\bdT_k \succeq \bdzro\comma\ \trace(\bdT_k) \le P_k\comma\ \forall k \in \clK} \ \sum_{k=1}^K \trace \left( \bdM_k^{(n)} \bdT_k \right). \label{subproblem_feasible_direction_Tk}
\end{align}
The closed-form solution to the problem in \eqref{subproblem_feasible_direction_Tk} can be written as \cite{Horn2013MatrixAnalysis}
\begin{align}
	\tdbdT_k^{(n+1)} = P_k \cdot \bdm_{k,n} \bdm_{k,n}^{H}\comma\ \forall k \in \clK\comma
\end{align}
where $\bdm_{k,n} \in \Complex{\Sone_k}{1}$ is the unit-norm eigenvector of $\bdM_k^{(n)}$ corresponding to its maximum eigenvalue. In terms of the CG method, $\{\bdT_k^{(n+1)}\}_{k=1}^K$ is given by
\begin{align}
	\bdT_k^{(n+1)} = \bdT_k^{(n)} + \alpha_k^{(n)} \left( \tdbdT_k^{(n+1)} - \bdT_k^{(n)} \right)\comma\ \forall k \in \clK\comma \label{Tkn+1_add_feasible_direction}
\end{align}
where $\alpha_k^{(n)} \in (0,1]$ is the stepsize of UT $k$ in the $n$th iteration. Th values of stepsize can be determined by the line search method \cite{Bertsekas2015ConvexAlgorithm,BoydConvexOptimization}. Once the optimal lower-dimensional $\{\bdT_k\}_{k=1}^K$ are obtained, the optimal $\{\bdQ_k\}_{k=1}^K$ can be obtained immediately by using \Cref{Relation_Qk_Tk_optimal}. The CG method for UL transmit design is summarized in \Cref{algorithm_UL_sum_rate_Tk}.
\begin{algorithm}[!t]
	\caption{CG method for UL transmit design.}
	\label{algorithm_UL_sum_rate_Tk}
	\begin{algorithmic}[1]
		\REQUIRE Initialize matrices $\bdT_k^{(0)} = \frac{P_k}{\Sone_k} \cdot \bdI$, $\forall k \in \clK$, and iteration index $n = 0$.
		\ENSURE Transmit covariance matrices $\{\bdQ_k\}_{k=1}^K$.
		\WHILE 1
		\STATE Compute $\bdM_k^{(n)}$ and corresponding $\bdm_{k,n}$, $\forall k \in \clK$.
		\STATE Update $\{\bdT_k^{(n+1)}\}_{k=1}^K$ according to \eqref{Tkn+1_add_feasible_direction}.
		\IF{$n\ge \Niter - 1$ or $\sabs{ R_{\SUM}^{(n+1)} - R_{\SUM}^{(n)} } < \epsilon$}
		\STATE Set $\bdT_k: = \bdT_k^{(n+1)}$, $\forall k \in \clK$, \textbf{break}.
		\ELSE
		\STATE Set $n:=n+1$.
		\ENDIF
		\ENDWHILE
		\STATE Compute $\bdQ_k = \bdB_k \bdT_k \bdB_k^H$, $\forall k \in \clK$.
	\end{algorithmic}
\end{algorithm}

Due to the expectation in the ESR, the complicated sample average has to be used for the computation of $\bdM_k^{(n)}$ in \Cref{algorithm_UL_sum_rate_Tk}. Next, to avoid the  exhaustive sample average, the asymptotic expression of ESR is used for the transmit design, which has a lower computational complexity and can achieve near-optimal performance.

\subsection{Simplified CG Method for UL Transmit Design} \label{subsubsection_asyprog_Tk}
In this subsection, we utilize an asymptotic expression of the ESR and devise a simplified CG method to compute the transmit covariance matrices, which can approximate the ESR capacity. Compared with \Cref{algorithm_UL_sum_rate_Tk}, the time-consuming sample average when computing $\bdM_k^{(n)}$ is no longer required in the simplified CG method. 
 
Let $\bdomega_k \triangleq  \frac{\beta_k}{\kappa_k + 1} \bbE \left\{ \tdbdc_k \odot \tdbdc_k^* \right\} \in \Complex{\Sone_k}{1}$, and define $\brbdH_{\re} \in \Complex{M}{\tdS}$ as follows
\begin{align}
	\brbdH_{\re} = \left[ \sqrt{\frac{\kappa_1 \beta_1}{\kappa_1+1}} \bdg_1 \bdc_{1,0}^H \ \cdots \ \sqrt{\frac{\kappa_K \beta_K}{\kappa_K+1}} \bdg_K \bdc_{K,0}^H \right]\comma
\end{align}
with $\tdS = \sum_{k=1}^{K} \tdS_k$.
An asymptotic expression, a.k.a. the deterministic equivalent, of the ESR can be written as \cite{Wen2011SumRateMIMORician,Lu2016FreeDeterministic,KaizheXu2021OnSumRate,LiYou2021EnergyRIS}
\begin{align}
	R_{\SUM}\rightarrow \udR_{\SUM} ={}& \log \det \left( \bdI +\bdXi \bdT \right) + \log \det \left( \bdI + \bdPhi_{\rR} \right) \notag \\
	&\quad - \sum_{k=1}^{K} \gamma_k \bdomega_k^T \bdpsi_k\comma \label{DE_sum_rate}
\end{align}
where $\bdT = \diag( \bdT_1,\dots,\bdT_K ) \in \Complex{\tdS}{\tdS}$, $\bdpsi_k = [\psi_{k,1} \ \cdots \ \psi_{k,\Sone_k}]^T \in \Real{\Sone_k}{1}$. In \eqref{DE_sum_rate}, $\{(\gamma_k, \bdpsi_k)\}_{k=1}^K$ is the unique solution of the following equations
\begin{subequations} \label{gamma_psi_equations_DE}
	\begin{align}
		\gamma_k & = \bdg_k^H \Xinv{\bdI+\bdPsi} \bdg_k\comma                                          \\
		\bdpsi_k & = \diag\left( \bdT_k  \left\langle \Xinv{\bdI + \bdXi \bdT } \right\rangle_k \right)\comma
	\end{align}
\end{subequations}
$k\in\clK$, where $\langle \cdot \rangle_k$ means the operation of taking the $k$th sub-block along the diagonal of the matrix argument. In addition, $\bdXi \in \Complex{\tdS}{\tdS}$ and $\bdPsi \in \Complex{M}{M}$ in \eqref{gamma_psi_equations_DE} depend on $\{(\gamma_k, \bdpsi_k)\}_{k=1}^K$ by
\begin{subequations} \label{DE_Matrix_Xi_Psi}
	\begin{align}
		\bdXi  & = \bdPhi_{\rT} + \brbdH_{\re}^H \Xinv{\bdI+\bdPhi_{\rR}} \brbdH_{\re}\comma \label{DE_Matrix_Xi}            \\
		\bdPsi & = \bdPhi_{\rR} + \brbdH_{\re} \bdT \Xinv{\bdI + \bdPhi_{\rT} \bdT } \brbdH_{\re}^H\comma \label{DE_Matrix_Psi}
	\end{align}
\end{subequations}
respectively, where $\bdPhi_{\rT}$ and $\bdPhi_{\rR}$ are given by
\begin{subequations}
\begin{align}
	\bdPhi_{\rT} &= \diag\left( \bdPhi_{\rT,1},\dots,\bdPhi_{\rT,K} \right) \comma \\
	\bdPhi_{\rR} &= \sum_{k=1}^{K} \bdPhi_{\rR,k}\comma
\end{align}
\end{subequations}
respectively, with $\bdPhi_{\rT,k} = \gamma_k \cdot \diag(\bdomega_k) \in \Complex{\Sone_k}{\Sone_k}$ and $\bdPhi_{\rR,k} = \bdomega_k^T \bdpsi_k \cdot \bdg_k \bdg_k^H \in \Complex{M}{M}$.
It has been claimed that the unique solution $\{(\gamma_k,\bdpsi_k)\}_{k=1}^K$ to the equations in \eqref{gamma_psi_equations_DE} can be obtained by performing a fixed-point iterative procedure until convergence \cite{Wen2011SumRateMIMORician,Lu2016FreeDeterministic,KaizheXu2021OnSumRate,LiYou2021EnergyRIS}. Then, the asymptotic approximation of the problem in \eqref{Problem_UL_sum_rate_Tk} can be formulated as follows
\begin{align}	\label{Problem_UL_asy_sum_rate_Tk}
	\udC = \max_{\bdT_k \succeq \bdzro\comma\ \trace(\bdT_k) \le P_k\comma \ \forall k}\ \udR_{\SUM}.
\end{align}
The gradient of $\udR_{\SUM}$ with respect to $\bdT_k$ is given by \cite{Wen2011SumRateMIMORician,Lu2016FreeDeterministic,KaizheXu2021OnSumRate,LiYou2021EnergyRIS}
\begin{align}
	\udbdM_k = \nabla_{\bdT_k} \udR_{\SUM}  = \left\langle \Xinv{ \bdI + \bdXi \bdT } \bdXi \right\rangle_k. \label{Vsum_gradient_Tk}
\end{align}
For given $\{\bdT_k^{(n)}\}_{k=1}^K$, a feasible solution of the problem in \eqref{Problem_UL_asy_sum_rate_Tk} in the $n$th iteration can be computed by solving the following subproblem
\begin{align}
	\max_{\bdT_k \succeq \bdzro\comma\ \trace(\bdT_k) \le P_k\comma\ \forall k \in \clK} \ \sum_{k=1}^K \trace \left( \udbdM_k^{(n)} \bdT_k \right). \label{subproblem_feasible_direction_Tk_DE}
\end{align}
The closed-form solution to the problem in \eqref{subproblem_feasible_direction_Tk_DE} can be written as \cite{Horn2013MatrixAnalysis}
\begin{align}
	\tilde{\udbdT}_k^{(n+1)} = P_k \cdot \udbdm_{k,n} \udbdm_{k,n}^{H}\comma\ \forall k \in \clK\comma
\end{align}
where $\udbdm_{k,n} \in \Complex{\Sone_k}{1}$ is the unit-norm eigenvector of $\udbdM_k^{(n)}$ corresponding to its maximum eigenvalue. Then, $\{\bdT_k^{(n+1)}\}_{k=1}^K$ is given by
\begin{align}
	\bdT_k^{(n+1)} = \bdT_k^{(n)} + \udalpha_k^{(n)} \left( \tilde{\udbdT}_k^{(n+1)} - \bdT_k^{(n)} \right)\comma\ \forall k \in \clK\comma \label{Tkn+1_add_feasible_direction_DE}
\end{align}
where $\udalpha_k^{(n)} \in (0,1]$ is the stepsize of the UT $k$ in the $n$th iteration. After $\{\bdT_k^{(n)}\}_{k=1}^K$ converges, $\{\bdQ_k\}_{k=1}^K$ can be obtained with \eqref{Relation_Qk_Tk_optimal}. The simplified CG method for UL transmit design with asymptotic ESR is shown in \Cref{algorithm_UL_sum_rate_Tk_DE}.
\begin{algorithm}[!t]
	\caption{Simplified CG method for UL transmit design.}
	\label{algorithm_UL_sum_rate_Tk_DE}
	\begin{algorithmic}[1]
		\REQUIRE Initialize matrices $\bdT_k^{(0)} = \frac{P_k}{\Sone_k} \cdot \bdI$, $\forall k \in \clK$, and iteration index $n = 0$.
		\ENSURE Transmit covariance matrices $\{\bdQ_k\}_{k=1}^K$.
		\WHILE 1
		\STATE Calculate $\udbdM_k^{(n)}$ and corresponding $\udbdm_{k,n}$, $\forall k \in \clK$.
		\STATE Update $\{\bdT_k^{(n+1)}\}_{k=1}^K$ according to \eqref{Tkn+1_add_feasible_direction_DE}.
		\IF{$n\ge \Niter - 1$ or $\sabs{ \udR_{\SUM}^{(n+1)} - \udR_{\SUM}^{(n)} } < \epsilon$}
		\STATE Set $\bdT_k: = \bdT_k^{(n+1)}$, $\forall k \in \clK$, \textbf{break}.
		\ELSE
		\STATE Set $n:=n+1$.
		\ENDIF
		\ENDWHILE
		\STATE Compute $\bdQ_k = \bdB_k \bdT_k \bdB_k^H$, $\forall k \in \clK$.
	\end{algorithmic}
\end{algorithm}
In terms of the number of multiplication operations, the computational complexity of \Cref{algorithm_UL_sum_rate_Tk_DE} is $\clO(\tdS^3 + K^3 + K^2 M)$.

\section{Simulation Results} \label{section_simulation}
\begin{table}[!t]
	\centering
	\footnotesize
	\renewcommand\arraystretch{1.2}
	\captionof{table}{Simulation Parameters}
	\label{table_simulation}
	\begin{tabular}{lc}
		\toprule
		Parameters                                                   	& Values                                                         \\
		\midrule
		Earth radius $R_e$                                         & $6378$ km                                                      \\
		Orbit altitude $H$                                          & $1000$ km                                                      \\
		Central frequency $f_c$                                & $2$ GHz                                                        \\
		Bandwidth $B$                                                & $20$ MHz                                                       \\
		Noise temperature $T_\rn$                         & $273$ K                                                        \\
		Number of antennas at satellite $\Mx\times \My$    & $12\times 12$                                    \\
		Number of antennas at UTs $\Nx\times \Ny$	  & $6\times 6$  \\
		Antenna spacing at satellite $d_{\rx}(d_{\ry})$ & $\lambda$ \\
		Antenna spacing at UTs $d_{\rx'}(d_{\ry'})$  & $\frac{\lambda}{2}$ \\
		Antenna gain at satellite $G_{\sat}$                          & $7$ dBi                                                \\
		Antenna gain at UTs  $G_{\ut}$ & $0$ dBi  \\
		Maximum nadir angle $\theta_{\max}^{\rz}$                         & $\xdeg{30}$                                                    \\
		Number of UTs $K$                                            & $100$                                                          \\
		Transmit power per UT                                        & $20$ dBm -- $40$ dBm                                          \\
		\bottomrule
	\end{tabular}
\end{table}

In this section, we provide the simulation results to verify the performance of the proposed UL transmit designs in massive MIMO LEO SATCOM.
The simulation parameters are summarized in \Cref{table_simulation}.
We denote the maximum nadir angle of the UTs as $\theta_{\max}^{\rz}$. The space angle pair $\bdzeta_k = (\zeta_k^{\rx},\zeta_k^{\ry})$ is generated according to the uniform distribution in the circle $\{(x,y)|x^2 + y^2 \le \sin^2 \theta_{\max}^{\rz}\}$.
The elevation angle of UT $k$ in \Cref{fig_uplink_UPA} is given by $v_k = \cos^{-1} \left( \frac{R_s}{R_e} \sin \theta_k^{\rz} \right)$, where $R_e$ is the earth radius, $R_s = R_e + H$ is the orbit radius  \cite{Maral2020SatelliteCommunications}.
The distance between the satellite and UT $k$ in \Cref{fig_uplink_UPA} is given by $D_k = \sqrt{R_e^2 \sin^2 v_k + H^2 + 2 H R_e} - R_e \sin v_k$ \cite{3GPP_NonTerrestrial}.
The per-antenna gains of the UPAs at the satellite and the UT sides are denoted as $G_{\sat}$ and $G_{\ut}$, respectively. 
The pathloss and shadow fading are computed according to the model parameters in \cite[Section 6]{3GPP_NonTerrestrial}, and the ionospheric loss is set as $2$ dB approximately \cite[Section 6]{3GPP_NonTerrestrial}.
For simplicity, we assume that each UT's UPA is placed horizontally, which implies that the paired AoDs $\bdvphi_{k,0}=(\vphi_{k,0}^{\rx'},\vphi_{k,0}^{\rz'})$ associated with the LoS path of UT$k$'s channel satisfies $\vphi_{k,0}^{\rz'}= \xdeg{90} - v_k$.
In order to obtain the covariance matrices $\bdSigma_k$'s, the orthogonal eigenvectors $\{\bdu_{k,i}\}_{i=1}^{S_k}$ are constructed by $\bdu_{k,i} = \bda_{\Nx}(\phi_k^{\rx'} + 2i/\Nx) \otimes \bda_{\Ny}(\phi_k^{\ry'} + 2i/\Ny)$ with $\phi_k^{\rx'} = \sin\vphi_{k,0}^{\rz'} \cos \vphi_{k,0}^{\rx'}$ and $\phi_k^{\ry'} = \sin \vphi_{k,0}^{\rz'} \sin \vphi_{k,0}^{\rx'}$, 
while the eigenvalues $\{\lambda_{k,i}\}_{i=1}^{S_k}$ are first randomly chosen according to the uniform distribution $\mathrm{U}(0,1)$ and then re-scaled such that $\sum_{i=1}^{S_k} \lambda_{k,i} = 1$. For simplicity, the rank of $\bdSigma_k$ is set as $S_k = 1$ in the following simulations, $\forall k \in \clK$.
The noise variance is given by $\sigma_k^2 = k_\rB T_\rn B$, where $k_\rB = 1.38 \times 10^{-23} \text{ J} \cdot \text{K}^{-1}$ is the Boltzmann constant, $T_{\rn}$ is the noise temperature and $B$ is the system bandwidth.

\begin{figure}[!t]
	\centering
	\includegraphics[width=0.5\textwidth]{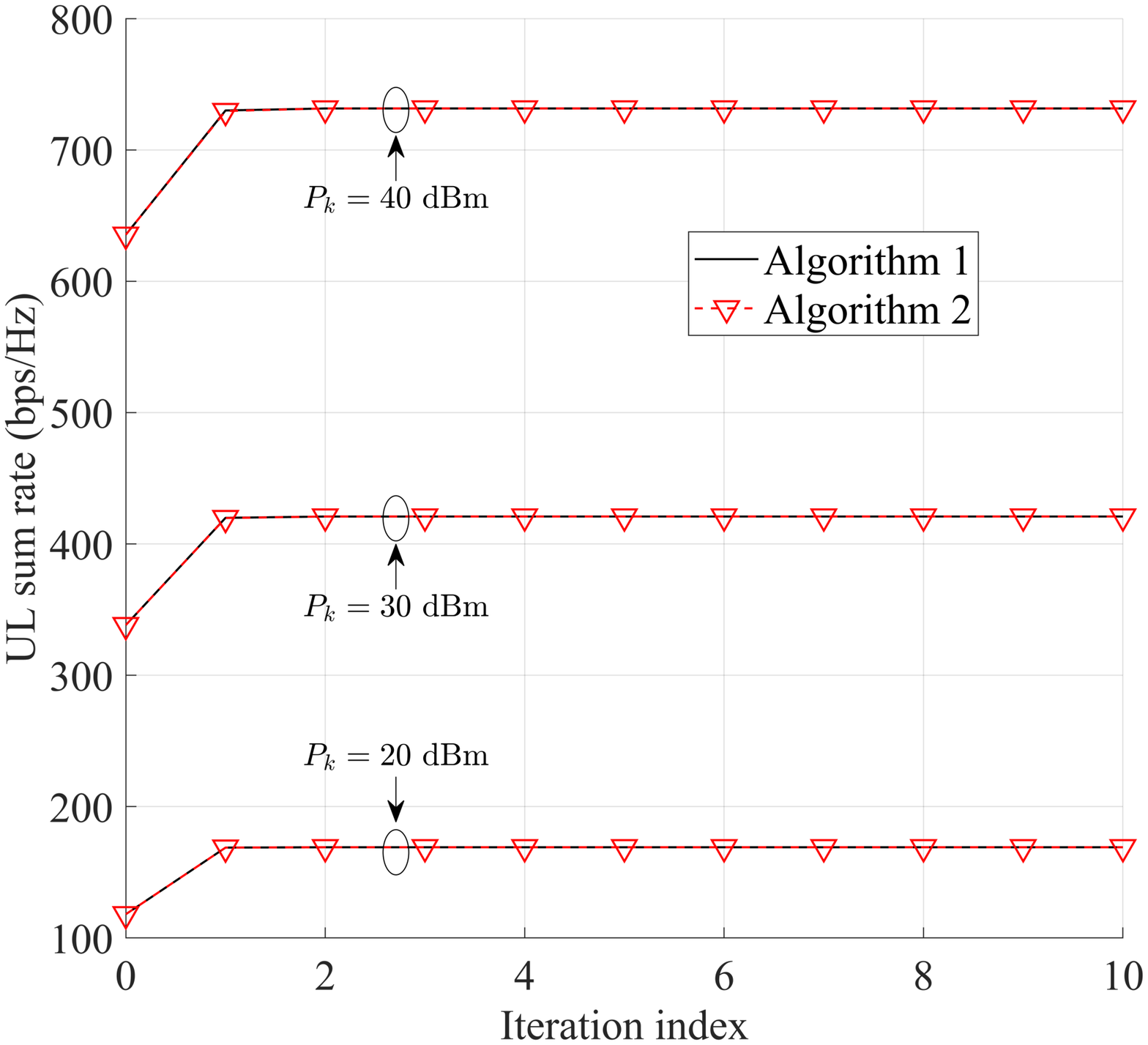}
	\caption{Convergence of \Cref{algorithm_UL_sum_rate_Tk,algorithm_UL_sum_rate_Tk_DE} at different transmit power.}
	\label{fig_convergence}
\end{figure}

\begin{figure}[!t]
	\centering
	\includegraphics[width=0.5\textwidth]{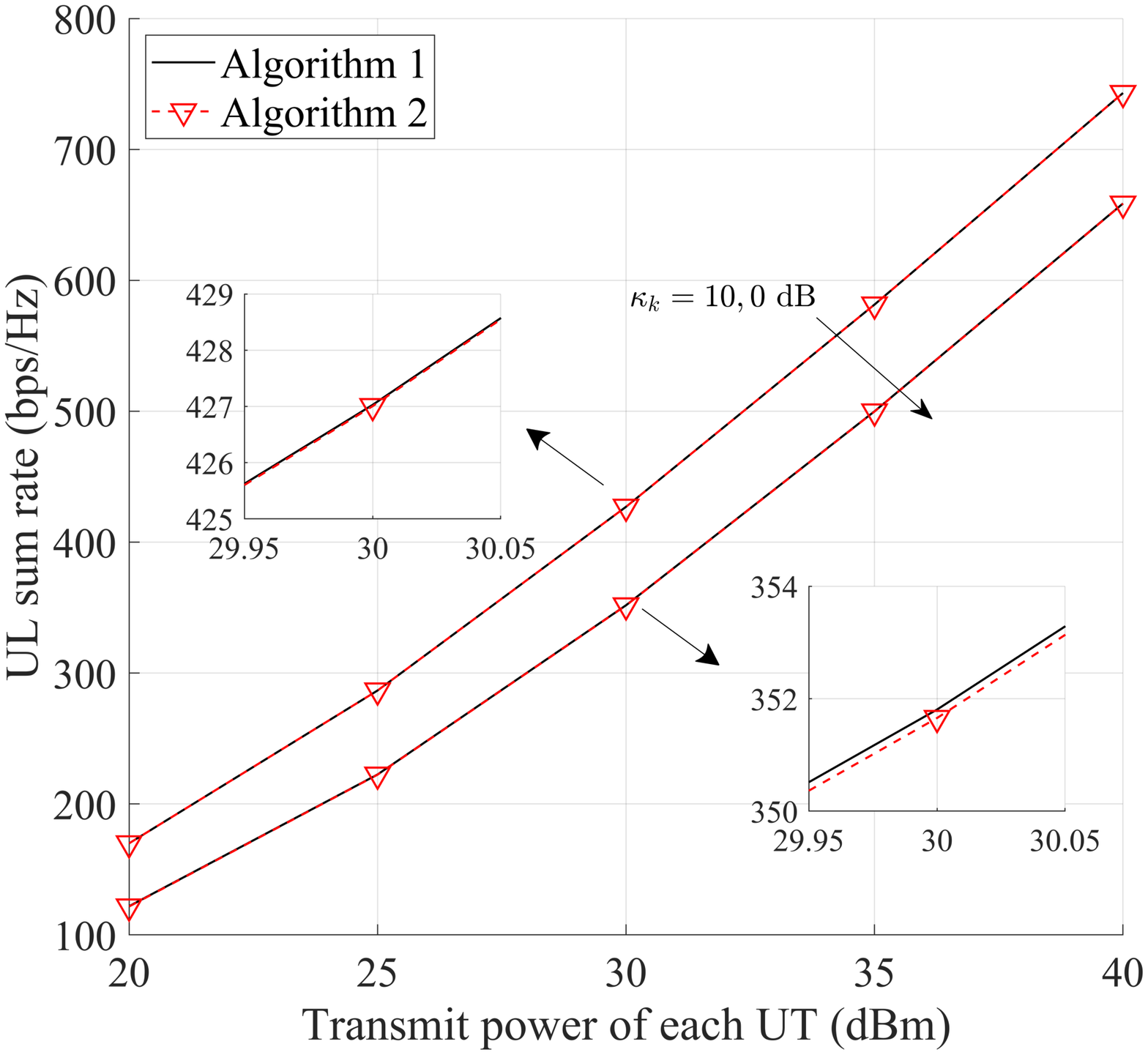}
	\caption{Performance of \Cref{algorithm_UL_sum_rate_Tk,algorithm_UL_sum_rate_Tk_DE} for different Rician factors.}
	\label{fig_performance_diffRice}
\end{figure}

In \Cref{fig_convergence}, the convergence performance of \Cref{algorithm_UL_sum_rate_Tk,algorithm_UL_sum_rate_Tk_DE} is depicted.  The transmit power $P_k$ of each UT $k$ takes different values including $20$ dBm, $30$ dBm, and $40$ dBm.
The Rician factor $\kappa_k$ of each UT $k$ is set as $\kappa_k = 10$ dB, $\forall k \in \clK$. To solve the fixed-point equations in \eqref{gamma_psi_equations_DE}, the number of iterations, $N_{\rf}$, is set as $N_{\rf} = 10$ hereafter.
As we can see, both \Cref{algorithm_UL_sum_rate_Tk,algorithm_UL_sum_rate_Tk_DE} can converge to the optimal values within a very small number of iterations.

\Cref{fig_performance_diffRice} shows the performance of \Cref{algorithm_UL_sum_rate_Tk,algorithm_UL_sum_rate_Tk_DE} for different values of Rician factors. It can be observed that the performance loss between \Cref{algorithm_UL_sum_rate_Tk,algorithm_UL_sum_rate_Tk_DE} is negligible for both high and low Rician factor cases.
In addition, we notice that the increase of Rician factors may bring the improvement of UL ESR capacity in massive MIMO LEO SATCOM, which means that the LoS components in LEO satellite channels may be more favorable to UL transmission. 

\begin{figure}[!t]
	\centering
	\includegraphics[width=0.5\textwidth]{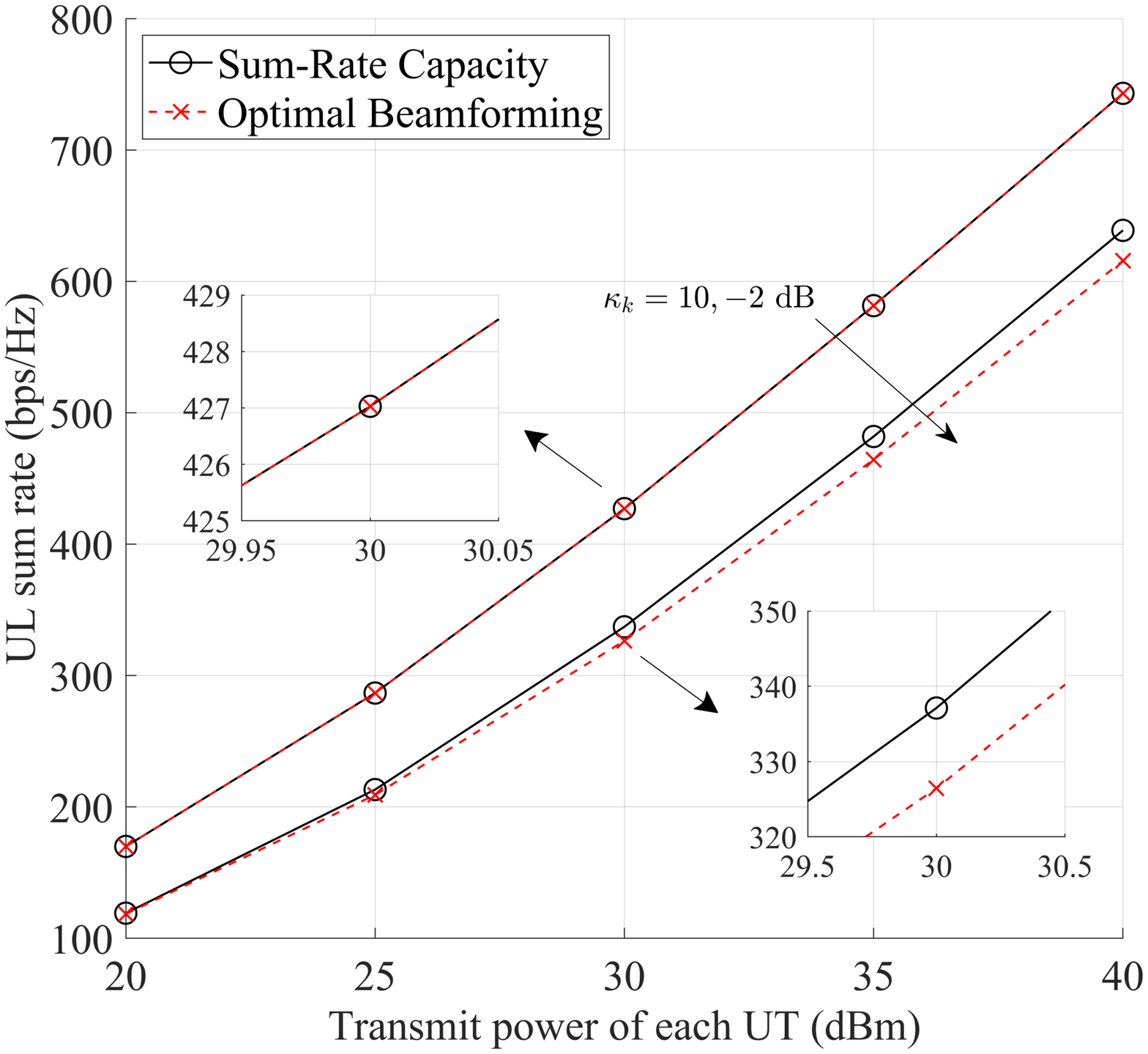}
	\caption{Comparison of ESR capacity and ESR with optimal beamforming.}
	\label{fig_performance_capacity_beamforming}
\end{figure}

\begin{figure}[!t]
	\centering
	\includegraphics[width=0.5\textwidth]{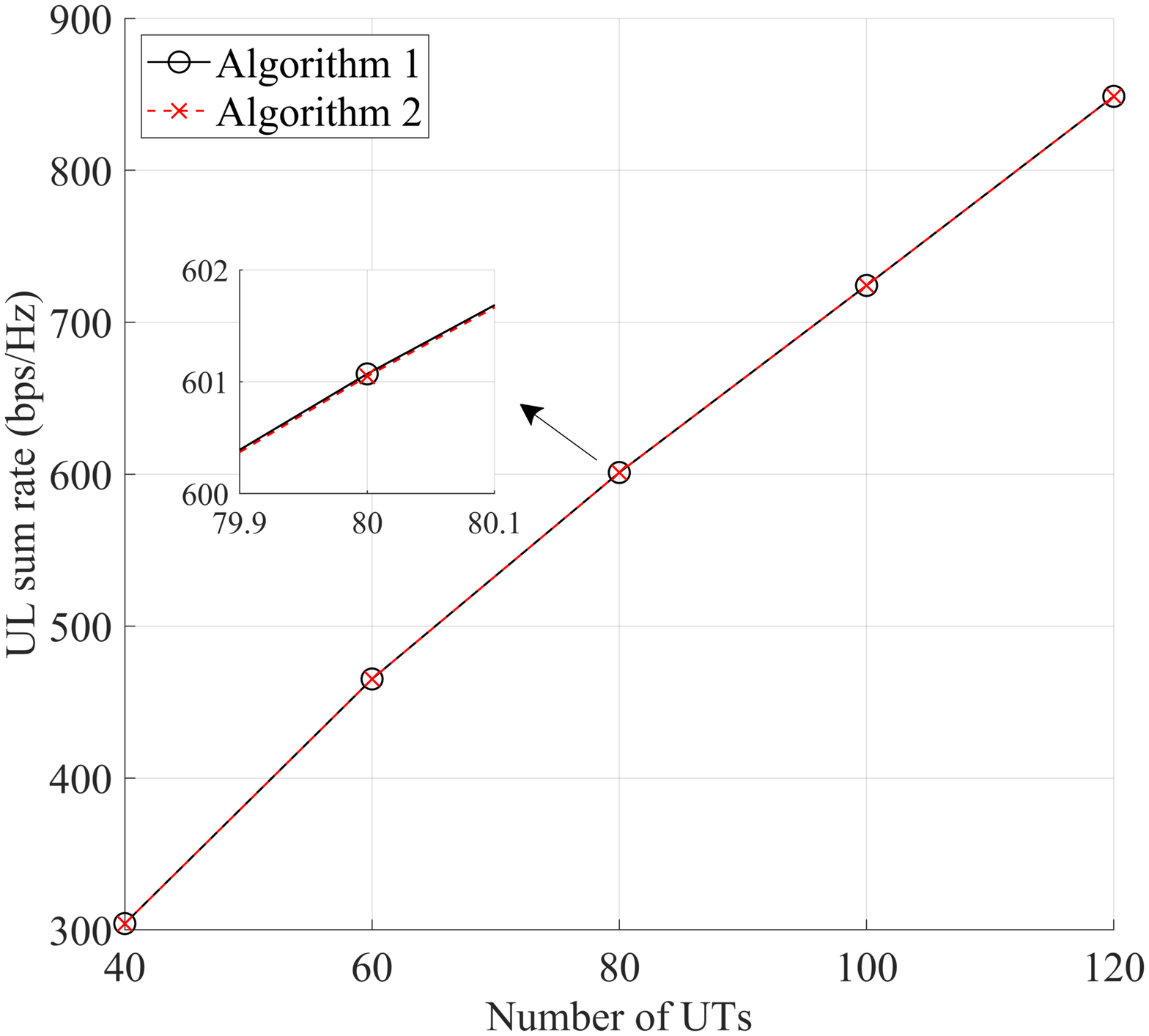}
	\caption{Performance of \Cref{algorithm_UL_sum_rate_Tk,algorithm_UL_sum_rate_Tk_DE} for different numbers of UTs.}
	\label{fig_performance_diffK}
\end{figure}

In \Cref{fig_performance_capacity_beamforming}, the ESR capacity and the optimal ESR attained by single data stream transmission from each UT to the satellite is compared. 
The ESR capacity is calculated by using \Cref{algorithm_UL_sum_rate_Tk}. Meanwhile, the optimal ESR with single data stream transmission from each UT to the satellite can be attained by using the optimal beamforming vectors of UTs denoted as $\{\bdB_k \bdw_{k}^{\opt}\}_{k=1}^K$ with 
$\{\bdw_{k}^{\opt}\}_{k=1}^K = \arg \max_{\snorm{\bdw_{k}}=1\comma\ \forall k\in\clK} \bbE \{ \log \det ( \bdI_M + \sum_{k=1}^K \frac{P_k}{\sigma^2}  \sabs{\bdc_k^H \bdw_{k}}^2 \cdot \bdg_k \bdg_k^H ) \}$, and this lower-dimensional vector optimization problem can be solved based on the CG method, which is omitted for conciseness.
It can be observed that when the Rician factors are relatively large, the ESR with optimal beamforming can be very close to the ESR capacity. Meanwhile, if the Rician factors are reduced to a relatively low level, which is likely to occur in some scenarios where the LoS components of UTs' channels suffer from some degree of blockage, the performance gap between the ESR capacity and the ESR with optimal beamforming will be non-negligible.

In \Cref{fig_performance_diffK}, the performance of \Cref{algorithm_UL_sum_rate_Tk,algorithm_UL_sum_rate_Tk_DE} for different numbers of UTs is depicted, in which the Rician factors are set as $\kappa_k = 10$ dB, $\forall k \in \clK$.
It can be seen that the ESR attained by \Cref{algorithm_UL_sum_rate_Tk_DE} is still tight to the ESR capacity achieved by using \Cref{algorithm_UL_sum_rate_Tk} for different numbers of UTs. In addition, as the number of UTs increases, the ESR capacity can be significantly improved in massive MIMO LEO SATCOM.

\section{Conclusion} \label{section_conclusion}

We have investigated the ESR capacity achieving UL transmit design with long-term sCSIT in massive MIMO LEO SATCOM systems. The UL massive MIMO LEO satellite channel model is established, where the satellite and UTs are equipped with UPAs. We prove that the rank of each UT's optimal transmit covariance matrix does not exceed that of its channel correlation matrix at the UT side. This reveals the maximum number of independent data streams transmitted from each multi-antenna UT to the satellite. We then prove that the transmit covariance matrix design can be transformed into the lower-dimensional matrix design without any loss of optimality. We obtain a necessary and sufficient condition when single data stream transmission from each UT to the satellite can achieve the ESR capacity.
A CG method is developed to compute the ESR capacity achieving transmit covariance matrices with guaranteed convergence. Further, to avoid the exhaustive sample average, we resort to an asymptotic expression of the ESR and devise a simplified CG method to compute the transmit covariance matrices which can approximate the ESR capacity. The effectiveness of the proposed approaches is verified in the simulation results.

\appendices

\section{Proof of \Cref{Proposition_1}} \label[secinapp]{appendix_Proposition_1_Proof}
The Lagrangian to the problem in \eqref{UL_Capacity_dQd} is given by
\begin{align}
	\clL_{\rQ} ={} & \bbE \left\{ \log \det \left( \bdI_M + \xinv{\sigma^2} \sum_{k=1}^{K} \bdd_k^H \bdQ_k \bdd_k \cdot \bdg_k \bdg_k^H \right) \right\}\notag \\
	             & \quad  + \sum_{k=1}^K \trace (\bdS_k \bdQ_k)  - \sum_{k=1}^K \mu_k \left( \trace(\bdQ_k) - P_k \right)\comma
\end{align}
where $\mu_k \ge 0$ and $\bdS_k \succeq \bdzro$ are Lagrange multipliers associated with the constraints $\trace(\bdQ_k) \le P_k$ and $\bdQ_k \succeq \bdzro$, respectively. At the optimum to the problem in \eqref{UL_Capacity_dQd}, the gradient of $\clL_{\rQ}$ with respect to $\bdQ_k$ will vanish, i.e.,
\begin{equation}
	\nabla_{\bdQ_k} \clL_{\rQ} = \bbE \left\{  W_k \bdd_k \bdd_k^H \right\} - \mu_k \bdI_N + \bdS_k = \bdzro\comma \label{sum_rate_capacity_Lagrangian_grad}
\end{equation}
where $W_k = \frac{ \xinv{\sigma^2} \bdg_k^H \bdA_k^{-1} \bdg_k }{ 1 + \xinv{\sigma^2} \bdg_k^H \bdA_k^{-1} \bdg_k \cdot \bdd_k^H \bdQ_k \bdd_k }$  with $\bdA_k =  \bdI_M + \xinv{\sigma^2} \sum_{i \ne k} \bdd_i^H \bdQ_i \bdd_i \cdot \bdg_i \bdg_i^H$.
By multiplying $\bdQ_k$ with the equation in \eqref{sum_rate_capacity_Lagrangian_grad} from the right side, it yields
\begin{equation}
	\bbE \left\{ W_k \bdd_k \bdd_k^H \right\} \bdQ_k = \mu_k \bdQ_k\comma \label{sum_rate_capacity_relation}
\end{equation}
where the complementary slackness condition $\bdS_k \bdQ_k = \bdzro$ in the Karush-Kuhn-Tucker (KKT) conditions is adopted \cite{BoydConvexOptimization}. 
We may now notice that the Lagrange multiplier $\mu_k$ must be strictly positive, $ \forall k \in \clK$. Otherwise,  the zero matrices $\{\bdQ_k= \bdzro\}_{k=1}^K$ will be an optimal solution to the problem in \eqref{UL_Capacity_dQd}, which is clearly not true. From the equality in \eqref{sum_rate_capacity_relation}, we can derive
\begin{align}
	\rank(\bdQ_k)  &= \rank \left( \bbE \left\{ W_k \bdd_k \bdd_k^H \right\} \bdQ_k \right) \notag \\
	&\le \rank \left( \bbE \left\{ W_k \bdd_k \bdd_k^H \right\} \right). \label{sum_rate_capacity_rank_Edd}
\end{align}
In addition, for every random vector $\bdd_k$, we have
\begin{equation}
	W_k \bdd_k \bdd_k^H  \stackpreceq{a} \xinv{\sigma^2} \bdg_k^H \bdA_k^{-1} \bdg_k \cdot \bdd_k \bdd_k^H\comma \label{sum_rate_capacity_dd}
\end{equation}
where (a) follows from $\bdd_k^H \bdQ_k \bdd_k \ge 0$. By taking the mathematical expectation for all random vector $\bdd_k$ in \eqref{sum_rate_capacity_dd}, we have
\begin{equation}
	\bbE \left\{ W_k \bdd_k \bdd_k^H \right\} \preceq \xinv{\sigma^2} \bbE \{ \bdg_k^H \bdA_k^{-1} \bdg_k \} \cdot \bdR_k^{\ut}\comma \label{sum_rate_capacity_expectation_dd}
\end{equation}
where $\bbE \{ \bdg_k^H \bdA_k^{-1} \bdg_k \} > 0$ and $\bdR_k^{\ut} = \bbE \{\bdd_k \bdd_k^H\}$.
By applying the inequality $\rank(\bdA) \ge \rank(\bdB)$ for $\bdA \succeq \bdB \succeq \bdzro$ \cite[Theorem 7.8]{Zhang2011MatrixTheory}, the relation in \eqref{sum_rate_capacity_expectation_dd} implies that
\begin{equation}
	\rank \left( \bbE \left\{ W_k \bdd_k \bdd_k^H \right\} \right) \le \rank \left( \bdR_k^{\ut} \right). \label{sum_rate_capacity_rank_final}
\end{equation}
After combining the results in \eqref{sum_rate_capacity_rank_Edd} and \eqref{sum_rate_capacity_rank_final}, we can complete the proof.

\section{Proof of \Cref{Proposition_2}} \label[secinapp]{appendix_Proposition_2_proof}
By utilizing $\bdd_k = \bdB_k \bdc_k$ in \eqref{dk_Bk_ck_equality}, the UL ESR capacity in \eqref{UL_Capacity_dQd} can be rewritten as
\begin{align}
	C={}&\max_{\bdQ_k \succeq \bdzro\comma\ \trace(\bdQ_k) \le P_k\comma\ \forall k}\notag \\
	& \bbE \left\{ \log \det \left( \bdI_M + \xinv{\sigma^2} \sum_{k=1}^K \bdc_k^H \bdB_k^H \bdQ_k \bdB_k \bdc_k \cdot \bdg_k \bdg_k^H \right) \right\}. \label{Rsum_Qk_Sk}
\end{align}
For any given positive semidefinite matrix $\bdT_k$, the linear matrix equation $\bdB_k^H \bdQ_k \bdB_k = \bdT_k$ always has a positive semidefinite solution $\bdQ_k$ if and only if $\bdT_k \bdB_k^{-} \bdB_k = \bdT_k$ \cite{GRO2000AXA=B,WenqianWu2021Shallow},
where $\bdB_k^{-} \in \Complex{\Sone_k}{M}$ is an arbitrary generalized inverse (g-inverse) of $\bdB_k$. The g-inverse $\bdA^{-}$ of $\bdA$ is defined as the matrix that satisfies the equation $\bdA \bdA^- \bdA = \bdA$. In terms of \cite[pp.~47]{Adi2003GeneralInverse}, one of the g-inverses of $\bdB_k$ is given by $\bdB_k^- = \Xginv{\bdB_k^H \bdB_k} \bdB_k^H = \bdB_k^H$ due to $\bdB_k^H \bdB_k = \bdI$.
Thus, we have
\begin{align} \label{Bk-Bk_Uk-Uk}
	\bdT_k \bdB_k^- \bdB_k = \bdT_k \bdB_k^H \bdB_k \stackeq{a} \bdT_k\comma
\end{align}
where (a) follows from $\bdB_k^H \bdB_k = \bdI$.
From the relation in \eqref{Bk-Bk_Uk-Uk}, we can conclude that for any given positive semidefinite $\bdT_k$, there always exists a positive semidefinite solution $\bdQ_k$ to the equation $\bdB_k^H \bdQ_k \bdB_k = \bdT_k$.

According to \cite{GRO2000AXA=B}, for a given positive semidefinite matrix $\bdT_k$, the positive semidefinite solution $\bdQ_k$ to the equation $\bdB_k^H \bdQ_k \bdB_k = \bdT_k$ can be represented by
\begin{equation}
	\begin{aligned}
		\bdQ_{N,k} = \bdB_k \bdT_k \bdB_k^H + \left( \bdI - \bdB_k \bdB_k^H \right) \bdN_k \left( \bdI - \bdB_k \bdB_k^H \right)\comma \label{Qk_g_inverse_Tk}
	\end{aligned}
\end{equation}
where $\bdN_k \in \Complex{N}{N}$ is an arbitrary positive semidefinite matrix. Then, the trace of matrix $\bdQ_{N,k}$ in \eqref{Qk_g_inverse_Tk} is given by
\begin{equation}
	\trace \left( \bdQ_{N,k} \right) = \trace \left( \bdT_k \right) + \Delta_k\comma
\end{equation}
where $\Delta_k = \trace \left( \left( \bdI - \bdB_k \bdB_k^H \right) \bdN_k \left( \bdI - \bdB_k \bdB_k^H \right) \right)$.
If $\Delta_k > 0$, we can always construct another matrix $\bdQ_{T,k} = \bdB_k \bdT_k \bdB_k^H$. Then, the matrix $\frac{\trace(\bdQ_{N,k})}{\trace(\bdT_{k})} \bdQ_{T,k}$ can attain a larger UL ESR $I_{\SUM}$ in \eqref{UL_Capacity} under the same power constraint with matrix $\bdQ_{N,k}$.
On the other hand, if $\Delta_k = 0$, which indicates that $\left( \bdI - \bdB_k \bdB_k^H \right) \bdN_k \left( \bdI - \bdB_k \bdB_k^H \right) = \bdzro$, then $\bdQ_{N,k}$ is exactly equal to $\bdB_k \bdT_k \bdB_k^H$.
Consequently, for a given matrix $\bdT_k$, we only need to consider the solutions $\bdQ_k$ to the linear matrix equation $\bdB_k^H \bdQ_k \bdB_k = \bdT_k$ with the following form
\begin{equation}
	\bdQ_k = \bdB_k \bdT_k \bdB_k^H. \label{Relation_Qk_Tk}
\end{equation}
Hence, the power constraint $\trace(\bdQ_k) \le P_k $ can be rewritten as
\begin{align}
	\trace(\bdQ_k) ={}& \trace(\bdB_k \bdT_k \bdB_k^H)\notag \\
	={}& \trace(\bdB_k^H \bdB_k \bdT_k) \notag \\
	\stackeq{a}{}& \trace(\bdT_k) \le P_k\comma
\end{align}
where (a) follows from $\bdB_k^H \bdB_k = \bdI$.
With the above statement, the problem in \eqref{UL_Capacity_dQd} can be equivalently transformed into the problem in \eqref{Problem_UL_sum_rate_Tk}. Furthermore, once the optimal solution $\{\bdT_k^{\star}\}_{k=1}^K$ to the problem in \eqref{Problem_UL_sum_rate_Tk} is obtained, the optimal matrices $\{\bdQ_k^{\star}\}_{k=1}^K$ to the problem in \eqref{UL_Capacity_dQd} can be directly derived with the aid of the relation in \eqref{Relation_Qk_Tk_optimal}.
This completes the proof.

\section{Proof of \Cref{Proposition_3}} \label[secinapp]{appendix_Prop_Tk_diagonal_proof}
In light of \eqref{dk0_uk0_pk0}, if $\bdxi_{k,0} = \bdzro$ holds for UT $k$, then $\udbdu_{k,0} = \bdu_{k,0} = \bdd_{k,0}$, $\tdS_k = S_k + 1$ and $\bdd_k = \bdB_k \bdc_k$ where $\bdB_k = [\bdd_{k,0}\ \bdU_k]$ and $\bdc_k$ is given by
\begin{align}
	\bdc_k =  \sqrt{\frac{\beta_k}{\kappa_k + 1}} \left[
	\begin{matrix}
		\sqrt{\kappa_k} \\
		\udtdbdc_{k}
	\end{matrix} \right]. \label{ck_proof_Rician}
\end{align}
Notice that $\udtdbdc_k$ is a CSCG random vector distributed as $\udtdbdc_k \sim \clCN\left(\bdzro,\diag(\bdlambda_k)\right)$.

Let $\bdT_{D,k}^{\circ}$ be the maximizer of $R_{\SUM} = \bbE \{ \log \det ( \bdI_M + \xinv{\sigma^2} \sum_{k=1}^K \bdc_k^H \bdT_k \bdc_k \cdot \bdg_k \bdg_k^H ) \}$ for any $\bdT_{D,k} \in \clD_k $, where $\clD_k = \{ \bdT_{k}| \bdT_k \succeq \bdzro\comma\ \trace(\bdT_k) \le P_k \text{ and }  \bdT_{k} \text{ is diagonal}  \}$.
For any $\bdT_{D,k} \in \clD_k$, the first order optimality condition must hold at $\bdT_{D,k}^{\circ}$ as follows
\begin{equation}
\trace \left(\left. \nabla_{\bdT_{k}}R_{\SUM} \right\rvert_{\bdT_{k} = \bdT_{D,k}^{\circ}} \left( \bdT_{D,k} - \bdT_{D,k}^{\circ} \right) \right) \le 0. \label{Proof_OptimalityCondition_Pk_trace}
\end{equation}
Let $G_k = \xinv{\sigma^2} \bdg_k^H \tdbdA_k^{-1} \bdg_k$ with $\tdbdA_k = \bdI_M + \xinv{\sigma^2} \sum_{i\ne k} \bdc_i^H \bdT_i \bdc_i \cdot \bdg_i \bdg_i^H$. Then, $\nabla_{\bdT_{k}}R_{\SUM}$ is given by
\begin{align}
	\nabla_{\bdT_{k}}R_{\SUM}  = \bbE \left\{ \frac{ G_k }{ 1 + G_k \bdc_k^H \bdT_{k} \bdc_k } \bdc_k \bdc_k^H \right\}.
\end{align}
Consequently, the inequality in \eqref{Proof_OptimalityCondition_Pk_trace} can be further written as
\begin{align} 
&\trace \left( \bbE \left\{ \frac{ G_k }{ 1 + G_k \bdc_k^H \bdT_{D,k}^{\circ} \bdc_k } \bdc_k \bdc_k^H \right\} \left( \bdT_{D,k} - \bdT_{D,k}^{\circ} \right) \right) \notag \\
={}& \bbE \left\{ \frac{ G_k \cdot \bdc_k^H ( \bdT_{D,k} - \bdT_{D,k}^{\circ} ) \bdc_k  }{ 1 + G_k \cdot \bdc_k^H \bdT_{D,k}^{\circ} \bdc_k } \right\} \le 0. \label{Proof_OptimalityCondition_Pk_scalar}
\end{align}
Then, we show that $\bdT_{D,k}^{\circ}$ is also the maximizer of $R_{\SUM}$ for all $\bdT_k \in \clN_k$, where $\clN_k = \{ \bdT_k| \bdT_k \succeq \bdzro\comma\ \trace(\bdT_k) \le P_k \}$.
To explain this, we separate $\bdT_k$ as $\bdT_k = \bdT_{G,k} + \bdT_{F,k}$, where $\bdT_{G,k}$ and $\bdT_{F,k}$ contain the diagonal and off-diagonal entries in $\bdT_k$, respectively. 
Then, we can obtain that
\begin{align}
& \bbE \left\{ \frac{ G_k \cdot \bdc_k^H ( \bdT_k - \bdT_{D,k}^{\circ} ) \bdc_k  }{ 1 + G_k \cdot \bdc_k^H \bdT_{D,k}^{\circ} \bdc_k } \right\} \notag \\
={}& \bbE \left\{ \frac{ G_k \cdot \bdc_k^H ( \bdT_{G,k} - \bdT_{D,k}^{\circ} ) \bdc_k  }{ 1 + G_k \cdot \bdc_k^H \bdT_{D,k}^{\circ} \bdc_k } \right\} \notag \\
&\quad + \bbE \left\{ \frac{ G_k \cdot \bdc_k^H \bdT_{F,k} \bdc_k  }{ 1 + G_k \cdot \bdc_k^H \bdT_{D,k}^{\circ} \bdc_k } \right\}. \label{Proof_OptimalityCondition_Pk_sum}
\end{align}
Notice that $\bdT_{G,k} \in \clD_k$ holds, because of $\trace(\bdT_k) = \trace(\bdT_{G,k}) \le P_k$. Combining \eqref{Proof_OptimalityCondition_Pk_scalar} and $\bdT_{G,k} \in \clD_k$, we can conclude that the first term in \eqref{Proof_OptimalityCondition_Pk_sum} is non-positive.
On the other hand, the second term in \eqref{Proof_OptimalityCondition_Pk_sum} can be rewritten as
\begin{align}
&\bbE \left\{ \frac{ G_k \cdot \bdc_k^H \bdT_{F,k} \bdc_k  }{ 1 + G_k \cdot \bdc_k^H \bdT_{D,k}^{\circ} \bdc_k } \right\} \notag \\
={}& \sum_{p=1}^{\Sone_k} \sum_{\substack{q=1, q\ne p}}^{\Sone_k} [\bdT_{F,k}]_{p,q} \cdot \bbE \left\{ J_{k} \left( [\bdc_k]_p, [\bdc_k]_q \right)  \right\} \comma \label{odd_function_ckp}
\end{align}
where $J_{k} \left( [\bdc_k]_p, [\bdc_k]_q \right)$ is defined as
\begin{align}
	J_{k} \left( [\bdc_k]_p, [\bdc_k]_q \right) = \frac{ G_k \cdot  [\bdc_k]_{p}^* [\bdc_{k}]_q  }{ 1 + G_k \cdot \sum_{i=1}^{\tdS_k} [\bdT_{D,k}^{\circ}]_{i,i} \abs{[\bdc_k]_{i}}^2 }\comma
\end{align}
with $[\bdc_k]_{p}$ denoting the $p$th element of $\bdc_k$. From \eqref{ck_proof_Rician}, $[\bdc_k]_p$ is either a constant or a CSCG random variable. Because $J_{k} \left( [\bdc_k]_p, [\bdc_k]_q \right) $ is an odd function with respect to $[\bdc_k]_p$ or $[\bdc_k]_q$, at least one of which is a CSCG random variable, $\bbE\{J_{k} \left( [\bdc_k]_p, [\bdc_k]_q \right) \} = 0$ must hold for any $ q \ne p$, so that \eqref{odd_function_ckp} is equal to zero. Therefore,  the following inequality holds for all $\bdT_k \in \clN_k$
\begin{align}
&\trace \left( \left. \nabla_{\bdT_k} R_{\SUM} \right\rvert_{\bdT_k = \bdT_{D,k}^{\circ}} \left( \bdT_k - \bdT_{D,k}^{\circ} \right) \right) \notag \\
={}& \bbE \left\{ \frac{ G_k \cdot \bdc_k^H ( \bdT_k - \bdT_{D,k}^{\circ} ) \bdc_k  }{ 1 + G_k \cdot \bdc_k^H \bdT_{D,k}^{\circ} \bdc_k } \right\} \notag \\
={}& \bbE \left\{ \frac{ G_k \cdot \bdc_k^H ( \bdT_{G,k} - \bdT_{D,k}^{\circ} ) \bdc_k  }{ 1 + G_k \cdot \bdc_k^H \bdT_{D,k}^{\circ} \bdc_k } \right\} \le 0. \label{Proof_OptimalityCondition_Pk_diagonal}
\end{align}
The condition in \eqref{Proof_OptimalityCondition_Pk_diagonal} means that $\bdT_k^{\star}$ will be a diagonal matrix under the condition $\bdxi_{k,0} = \bdzro$. 
This completes the proof.

\section{Proof of \Cref{Proposition_4}} \label[secinapp]{appendix_Prop_Tk_rankone_proof}
Let $\bdK_k \in \Complex{\tdS_k}{\tdS_k}$ be an arbitrary feasible matrix for the problem in \eqref{Problem_UL_sum_rate_Tk}, which means $\bdK_k \succeq \bdzro$ and $\trace(\bdK_k) \le P_k$. 
Define a set of lower-dimensional matrices as follows
\begin{align}
	\clW_k(\bdK_k) = \{(1-\rho)P_k \bdw_k \bdw_k^H + \rho \bdK_k|\rho \in[0,1]\}. \label{setdef_Wk_Kk}
\end{align} 
By taking all the possible feasible matrix $\bdK_k$, the union of the above sets will constitute the entire feasible matrix set $\{\bdT_k|\bdT_k \succeq \bdzro, \ \trace(\bdT_k)\le P_k\}$ of UT $k$ for the problem in \eqref{Problem_UL_sum_rate_Tk}. Therefore, for each UT $k$, the ESR maximization over the feasible matrix set $\{\bdT_k|\bdT_k \succeq \bdzro, \ \trace(\bdT_k)\le P_k\}$ can be transformed into that over the set $\clW_k(\bdK_k)$ for any feasible matrix $\bdK_k$.

For a given feasible matrix $\bdK_k$, define the ESR over the set $\clW_k(\bdK_k)$ as follows
\begin{align}
	F_k(\rho) ={}& \bbE \left\{ \log \det \left( \tdbdA_k + \xinv{\sigma^2} \bdc_k^H \left( P_k \bdw_k \bdw_k^H  \right.\right.\right. \notag \\
	& \quad \left.\left.\left. + \rho \left( \bdK_k - P_k \bdw_k\bdw_k^H \right)  \right) \bdc_k \cdot \bdg_k \bdg_k^H \right) \right\} \notag \\
	={}& \bbE \left\{ \log \det \left( \tdbdA_k \right) \right\} + \bbE \left\{ \log \left( 1 + G_k \cdot \bdc_k^H  \right.\right. \notag \\
	&\quad \left.\left. \left( P_k \bdw_k \bdw_k^H + \rho \left( \bdK_k - P_k \bdw_k\bdw_k^H \right)  \right) \bdc_k \right) \right\} \comma
\end{align}
where $ 0 \le \rho \le 1$, $\tdbdA_k = \bdI_M + \xinv{\sigma^2} \sum_{i\ne k} \bdc_i^H \bdT_i \bdc_i \cdot \bdg_i \bdg_i^H$ and $G_k = \xinv{\sigma^2} \bdg_k^H \tdbdA_k^{-1} \bdg_k$.
It can be seen that $F_k(\rho)$ is a concave function of $\rho$ \cite{BoydConvexOptimization}. A necessary condition for the optimality of  $\bdT_k^{\star} = P_k \bdw_k \bdw_k$ regarding the problem in \eqref{Problem_UL_sum_rate_Tk} is $\left. \frac{\dint  F_k(\rho)}{\dint \rho} \right\rvert_{\rho = 0} \le 0$ for any feasible matrix $\bdK_k$, which means that 
\begin{align}
	\max_{\bdK_k \succeq \bdzro\comma \ \trace(\bdK_k) \le P_k} \left( \left. \frac{\dint  F_k(\rho)}{\dint \rho} \right\rvert_{\rho = 0}  \right) \le 0. \label{suff_necc_condition_proof}
\end{align}
In addition, due to the concavity of $F_k(\rho) $, if \eqref{suff_necc_condition_proof} holds, the point $\rho = 0$ will be the global optimum of $F_k(\rho)$, for any feasible matrix $\bdK_k$. Furthermore, because $F_k(\rho)$ represents the ESR over the set $\clW_k(\bdK_k)$, by recalling the definition of $\clW_k(\bdK_k)$ in \eqref{setdef_Wk_Kk}, if $\rho = 0$ is the global optimum of $F_k(\rho)$ for any feasible matrix $\bdK_k$, $\bdT_k = P_k \bdw_k \bdw_k$ will be the optimal solution to the problem in \eqref{Problem_UL_sum_rate_Tk}. Therefore, \eqref{suff_necc_condition_proof} is a necessary and sufficient condition when $\bdT_k = P_k \bdw_k \bdw_k$ is optimal for the problem in \eqref{Problem_UL_sum_rate_Tk} .
In \eqref{suff_necc_condition_proof},  $\left. \frac{\dint  F_k(\rho)}{\dint \rho} \right\rvert_{\rho = 0} $ is given by
\begin{align}
	&\left. \frac{\dint  F_k(\rho)}{\dint \rho} \right\rvert_{\rho = 0} 
	= \bbE \left\{  \frac{ G_k \bdc_k^H \bdK_k \bdc_k - G_k P_k \sabs{\bdc_k^H \bdw_k}^2 }{ 1 + G_k P_k \sabs{\bdc_k^H \bdw_k}^2 } \right\}.
\end{align}

Let $\bdK_k = \sum_{i=1}^{\tdS_k} \xi_i \bdp_{i} \bdp_i^H$ denote the EVD of $\bdK_k$, where $\{\xi_i\}_{i=1}^{\tdS_k}$ and $\{\bdp_i\}_{i=1}^{\tdS_k}$ are the non-negative eigenvalues and the corresponding orthogonal eigenvectors, respectively. Then, the condition in \eqref{suff_necc_condition_proof} can be further written as
\begin{align}
	0 \ge{}& \max_{\xi_i, \ \bdp_{i}, \ \forall i } \bbE \left\{ \frac{ \sum_{i=1}^{\tdS_k} \xi_i G_k \bdc_k^H \bdp_{i} \bdp_i^H \bdc_k - G_k P_k \sabs{\bdc_k^H \bdw_k}^2 }{1+G_k P_k \sabs{\bdc_k^H \bdw_k}^2 } \right\} \notag \\
	={}& \max_{\xi_i, \ \bdp_{i}, \ \forall i } \sum_{i=1}^{\tdS_k} \xi_i \bdp_i^H \bbE \left\{ \frac{ G_k\bdc_k  \bdc_k^H }{1+G_k P_k \sabs{\bdc_k^H \bdw_k}^2 } \right\} \bdp_{i} \notag \\
	&\quad - \bbE \left\{ \frac{G_k P_k \sabs{\bdc_k^H \bdw_k}^2 }{1+G_k P_k \sabs{\bdc_k^H \bdw_k}^2 } \right\} \notag \\
	\stackeq{a}{}& P_k \left( \Upsilon_{\max} \left( \bbE \left\{ \frac{ G_k\bdc_k  \bdc_k^H }{1+G_k P_k \sabs{\bdc_k^H \bdw_k}^2 } \right\} \right) \right. \notag \\
	&\quad \left. - \bbE \left\{ \frac{G_k \sabs{\bdc_k^H \bdw_k}^2 }{1+G_k P_k \sabs{\bdc_k^H \bdw_k}^2 } \right\} \right)\comma \label{suff_necc_condition_ineq_1}
\end{align}
where (a) follows from $\bdp_i^H \bdX \bdp_{i} \le \Upsilon_{\max}\left( \bdX  \right)$ and $\sum_{i=1}^{\tdS_k} \xi_i \le P_k$.
Thus, the condition in \eqref{suff_necc_condition_ineq_1} can be written as 
\begin{align}
	 &\Upsilon_{\max} \left( \bbE \left\{ \frac{ G_k\bdc_k  \bdc_k^H }{1+G_k P_k \sabs{\bdc_k^H \bdw_k}^2 } \right\} \right) \notag \\
	  \le{}& \bbE \left\{ \frac{G_k \sabs{\bdc_k^H \bdw_k}^2 }{1+G_k P_k \sabs{\bdc_k^H \bdw_k}^2 } \right\}. \label{suff_necc_condition_ineq_2}
\end{align}
In addition, since $\bdw_k$ is a unit-norm vector, we have
\begin{align}
	&\bbE \left\{ \frac{G_k \sabs{\bdc_k^H \bdw_k}^2 }{1+G_k P_k \sabs{\bdc_k^H \bdw_k}^2 } \right\} \notag \\
	={}& \bdw_k^H  \bbE \left\{ \frac{G_k \bdc_k \bdc_k^H }{1+G_k P_k \sabs{\bdc_k^H \bdw_k}^2 } \right\} \bdw_k \notag \\
	\le{}& \Upsilon_{\max} \left( \bbE \left\{ \frac{ G_k\bdc_k  \bdc_k^H }{1+G_k P_k \sabs{\bdc_k^H \bdw_k}^2 } \right\} \right). \label{suff_necc_condition_ineq_3}
\end{align}
Consequently, by combining the results in \Cref{suff_necc_condition_ineq_2,suff_necc_condition_ineq_3}, the condition in \eqref{eigmax_wtk} can be obtained. This completes the proof.


\bibliographystyle{IEEEtran}
\bibliography{IEEEabrv,satellite_uplink_transceiver_lib}

\end{document}